\documentclass[prx,twocolumn,superscriptaddress,floatfix,nopacs]{revtex4}
\usepackage{graphicx,amsfonts,amssymb,amsmath,hyperref,enumerate}

\newif\ifhyper
\hypertrue
\ifhyper
\hypersetup{
   citecolor = {green},
   colorlinks = {true}, 
   urlcolor = {blue} 
}
\fi

\newcommand{\beq}{\begin{equation}}
\newcommand{\eeq}{\end{equation}}
\newcommand{\beqa}{\begin{eqnarray}}
\newcommand{\eeqa}{\end{eqnarray}}
\newcommand{\ket} [1] {\vert #1 \rangle}

\def\ket#1{\vert#1\rangle}

\def\Longarrow{\protect\@lra}
\def\@lra{\relbar\joinrel\relbar\joinrel\relbar\joinrel%
          \relbar\joinrel\rightarrow}

\begin{document}

\title{Tensor network simulation of QED on infinite lattices: \\Êlearning from $(1+1)d$, and prospects for $(2+1)d$}

\author{Kai Zapp}
\affiliation{Institute of Physics, Johannes Gutenberg University, 55099 Mainz, Germany}

\author{Rom\'an Or\'us}
\affiliation{Institute of Physics, Johannes Gutenberg University, 55099 Mainz, Germany}

\begin{abstract}
The simulation of lattice gauge theories with tensor network (TN) methods is becoming increasingly fruitful. The vision is that such methods will, eventually, be used to simulate theories in $(3+1)$ dimensions in regimes difficult for other methods. So far, however, TN methods have mostly simulated lattice gauge theories in $(1+1)$ dimensions. The aim of this paper is to explore the simulation of quantum electrodynamics (QED) on infinite lattices with TNs, i.e., fermionic matter fields coupled to a $U(1)$ gauge field, directly in the thermodynamic limit. With this idea in mind we first consider a gauge-invariant iDMRG simulation of the Schwinger model -i.e., QED in $(1+1)d$-. After giving a precise description of the numerical method, we benchmark our simulations by computing the substracted chiral condensate in the continuum, in good agreement with other approaches. Our simulations of the Schwinger model allow us to build intuition about how a simulation should proceed in $(2+1)$ dimensions. Based on this, we propose a variational ansatz using infinite Projected Entangled Pair States (PEPS) to describe the ground state of $(2+1)d$ QED. The ansatz includes $U(1)$ gauge symmetry at the level of the tensors, as well as fermionic (matter) and bosonic (gauge) degrees of freedom both at the physical and virtual levels. We argue that all the necessary ingredients for the simulation of $(2+1)d$ QED are, \emph{a priori}, already in place, paving the way for future upcoming results.  

\end{abstract}

\maketitle

\section{introduction}Ê
\label{sec1}

Gauge field theories \cite{QFT} are currently our deepest level of understanding of how fundamental interactions emerge from local symmetry principles. The standard modelÊ is a gauge theory, where different gauge symmetries orchestrate all known interactions except for gravity, which can be seen itself also as a gauge theory. The structure of gauge theories is so complex that, sometimes, it is wise to discretize them on a lattice in order to simulate their properties on a computer. Even if bumpy at its historical origins, the numerical simulation of lattice gauge theories \cite{lgt} has become one of the main tools to understand our universe. This is particularly true for quantum chromodynamics (QCD), the theory of strong interactions, where lattice simulations allowed to, e.g., understand the spectrum of hadrons observed in particle accelerators. 

Still, many questions concerning gauge theories remain open, and in particular for QCD. For instance, what is its phase diagram at finite fermionic density? Or what are the dynamical properties of the theory? Usual lattice gauge theory calculations, based mostly on quantum Monte Carlo, fail to answer such questions because of fundamental algorithmic limitations. Moreover, finite-size scaling of the results relies on accurate extrapolation laws to the thermodynamic limit which need to be somehow known beforehand. 

In this setting, tensor network (TN) numerical methods \cite{tn} have emerged as a promising alternative. In TN methods, the wavefunction of the system is decomposed into fundamental pieces, the tensors,  glued together by quantum entanglement according to some network pattern. Such methods rely on correctly reproducing the amount and structure of entanglement in the wavefunction being simulated. The methods usually target low-energy properties, but can also be adapted to compute dynamics. Moreover, one can simulate both bosons and fermions with essentially the same computational cost \cite{ftn}. And on top, gauge symmetries can  be implemented naturally in this framework \cite{gtn, gaugeMPS}. So all in all, TNs look like the natural option to describe the structure of quantum states present in lattice gauge theories. 

Our aim with this paper is to pave the way towards higher-dimensional numerical simulations of lattice gauge theories, in particular for $(2+1)d$ quantum electrodynamics (QED), i.e., the gauge theory for electromagnetism. In order to build intuition, we first do a detailed analysis of simulations of lattice QED in $(1+1)d$, the so-called \emph{Schwinger model} \cite{schwinger}, using gauge-invariant Matrix Product States (MPS) \cite{mps, gaugeMPS} and a gauge-invariant version of infinite Density Matrix Renormalization Group (iDMRG) \cite{dmrg, idmrg}. This allows us to foresee how a higher-dimensional simulation should proceed. For the higher-dimensional case we discuss briefly the lattice Hamiltonian, and give a proposal for a 2d TN ansatz based on infinite Projected Entangled Pair States (iPEPS) \cite{PEPS, iPEPS}. As we shall see, such an iPEPS implements naturally fermionic matter and $U(1)$ gauge bosons. Thinking in perspective, we argue that all the necessary ingredients for a TN simulation of QED in $(2+1)d$ are \emph{a priori} already there. 

Previous results on the TN simulation of lattice gauge theories include a number of works. $\mathbb{Z}_2$ lattice gauge theories in $(1+1)d$ have been considered with DMRG \cite{sugihara}.  For the Schwinger model, DMRG (without MPS formulation) was considered in several works \cite{dmrgOld}, whereas  MPS simulations have been done to compute the chiral condensate \cite{SchwingerDMRG} as well as thermal properties \cite{SchwingerThermal}, the mass spectrum \cite{SchwingerSpectrum}, the Schwinger effect \cite{SchwingerSchwinger}, the effect of truncation in the gauge variable \cite{SchwingerTruncation}, and the case of several fermionic flavours \cite{SchwingerMultiflavour}. Gauge invariance in the MPS of the Schwinger model was originally considered in Ref.\cite{gaugeMPS}, where the ground state was computed using the time-dependent variational principle (TDVP) \cite{tdvp}. Gauge-invariant MPS were used to compute the confining potential \cite{SchwingerPot}. A similar approach was used to analyze the scattering of two quasiparticles and the dynamical generation of entanglement \cite{SchwingerScat}. TN simulations have also been implemented recently for non-abelian lattice gauge theories in $(1+1)d$ \cite{nonabelianTNS}. For higher-dimensional systems, gauge-invariant TN ansatzs have also been proposed analytically \cite{gtn, u1peps, su2peps}. 

This paper is organized as follows: first, in Sec.\ref{sec2} we provide a detailed introduction to QED in $(1+1)$ dimensions (the Schwinger model) in the continuum as well as its discretized version on the lattice. In this section we provide also background on the so-called chiral condensate. Then, in Sec.\ref{sec3}Ê we revise the infinite-DMRG algorithm. We discuss the details of the variational optimization of MPS, with one-site and two-site updates in the thermodynamic limit. In Sec.\ref{sec4}Ê we explain how to do a gauge-invariant simulation of the Schwinger model using infinite DMRG. Numerical benchmarks for the chiral condensate in $(1+1)d$ are presented in Sec.\ref{sec5}, paying attention to the continuum limit extrapolation. Based on all this, in Sec.\ref{sec6}Ê we discuss the prospects for the simulation of QED in $(2+1)d$, where we consider the lattice formulation of the Hamiltonian as well as a possible TN ansatz for its ground state in terms of a 2d infinite PEPS. Finally, Sec.\ref{sec7} contains our conclusions. 

\section{QED in $(1+1)d$: the Schwinger model}
\label{sec2}

Let us now revise the basics of QED in $(1+1)$ dimensions, also called QED$_2$, or the Schwinger model \cite{schwinger}. We will refresh some of the properties of this theory defined in the continuum, as well as a possible formulation on the lattice, which
will be the starting point of our study with TN methods. Readers who are interested in a more detailed discussion of the model and its properties are referred to, e.g., Ref.~\cite{SchwingerDMRG}. 

\subsection{Continuum formulation}Ê

The massive Schwinger model is quantum electrodynamics in two space-time dimensions. Its Lagrangian density in the continuum reads
\begin{align}
\mathcal{L} = \overline{\psi}\left(i \partial_{\mu}\gamma^{\mu}  -m \right) \psi - \frac{1}{4} F_{\mu \nu} F^{\mu \nu} -  g \overline{\psi} A_{\mu} \gamma^{\mu} \psi,
\label{eg:Lagrangian}
\end{align}
where
\begin{align}
	F^{\mu \nu} = \partial^{\mu} A^{\nu} - \partial^{\nu} A^{\mu}.
\end{align}
The first term is the Dirac Lagrangian density for a free fermion and the second term corresponds to the field energy of the electric field (in $(1+1)d$ there is no ``room" for a magnetic field). The third term is the interaction between the matter field and the gauge field. It has the important feature that it arises from the constraints imposed by a \emph{local} gauge transformation. That means, its form is determined by demanding the invariance of the Lagrangian density under the transformation
\begin{align}
	\psi' = e^{i g \chi} \psi, \quad A'_{\mu} = A_{\mu} + \partial_{\mu} \chi,
\end{align}
where $\chi$ is an arbitrary real function of space and time \footnote{This is what is meant by \emph{local}: one can redefine the phase of the fermionic field locally at every point in space-time.}, i.e. $\chi=\chi\left(x,t\right)$.
The Schwinger model describes the interaction of one flavour mass-$m$ fermions $\psi$ with a $U(1)$ gauge field $A$, with coupling $g$. In $(1+1)d$ the Lorentz indices $\mu,\nu $ run from $0$ to $1$ (one direction for space, and one for time), and the gamma matrices satisfy the Clifford algebra
\begin{align}
	\left\lbrace \gamma^{\mu} , \gamma^{\nu}\right\rbrace = 2 g^{\mu \nu}, 
\end{align}
analogously to the $(3+1)d$ case. However, since there is no spin degree of freedom in one spatial dimension, these are $2\times 2$ matrices. Substituting the Lagrangian of the Schwinger model into Euler-Lagrange equations for the fields $\psi$ and $A$ results in
the equations of motion
\begin{align}
	&\gamma^{\mu} \left(i D_{\mu} - m\right)\psi = 0 ,
\end{align}
and
\begin{align}
	&\partial_{\mu} F^{\mu \nu} = g j^{\nu}, 
	\label{eq: Maxwell}
\end{align}
where $D_\mu \equiv \partial_\mu + igA_\mu$ is the gauge covariant derivative and $j^{\nu} \equiv \overline{\psi} \gamma^{\nu} \psi$ the vector current.
The theory is quantized using canononical quantization by imposing anti-commutation relations on the fermion fields
\beqa
	\left\lbrace \psi^{\dagger}\left(x,t\right) , \psi\left(x,t\right)\right\rbrace &=& \delta\left(x-y\right) \nonumber \\ 
        \left\lbrace \psi^{\dagger}\left(x,t\right) ,\psi^{\dagger}\left(x,t\right)\right\rbrace &=& \left\lbrace \vphantom{\psi^{\dagger}} \psi\left(x,t\right) ,\psi\left(x,t\right)\right\rbrace=0,
\eeqa
and by imposing commutation relations on the gauge fields
\begin{align}
	\left[E\left(x,t\right),A_{1}\left(y,t\right)\right] = i \delta\left(x-y\right),
\end{align}
where the electric field $E$ is defined by 
\begin{align}
	E = - F^{01} = F^{10}.
	\label{eq: E_def}
\end{align}
Using this definition of the electric field in Eq.(\ref{eq: Maxwell}), we get analogues to Maxwell's equations in $(1+1)d$:
\begin{align}
&	\frac{\partial E}{\partial x} = g j^{0} \equiv g \rho, \qquad \text{(Gauss' law)}\\  
	\notag
	- &\frac{\partial E}{\partial t} = g j^{1} \equiv g j. 
\end{align}
Since there is ``no space'' for magnetic fields in one spatial dimension, we only obtain the analogue of Gauss' law and an equation which describes the dynamics of the electric field. 


\subsection{Lattice formulation} 
Starting from the Hamiltonian density $\mathcal{H}$ in temporal gauge, $A_{0} = 0$,
\begin{align}
		\mathcal{H} = - i  \overline{\psi} \gamma^{1} \left(\partial_{1} -  i g A_{1}\right) \psi + m \overline{\psi} \psi + \frac{1}{2} E^2,
		\label{eq: Hamilton_density}
\end{align}
the model can be formulated on a spatial lattice using a Kogut-Susskind staggered formulation \cite{KogutSusskind}. The equivalent lattice Hamiltonian is 
\beqa
	H &= &- \frac{i}{2a} \sum_{n} \left(\phi^{\dag}_n e^{i \theta_n}\phi_{n+1} - h.c.\right) + m \sum_{n} \left(-1\right)^n \phi^{\dag}_n \phi_n \nonumber \\ 
	&+& \frac{a g^2}{2} \sum_{n} L^{2}_n. 
\eeqa
where $a$ denotes the lattice spacing. In this formulation the correspondence between the fermionic lattice field $\phi_n$ on site $n$ and the
continuum field $\psi$ is
\begin{align}
	\phi_n \leftrightarrow \begin{cases} \psi_{upper} \quad n \text{ even} \\ \psi_{lower} \quad n \text{ odd} \end{cases} , \qquad \psi = \begin{pmatrix}
		\psi_{upper} \\
		\psi_{lower}
	\end{pmatrix}.
\end{align}
The gauge variables $\theta_n$  live on the links between the sites $n$ and $n+1$, and are related to the vector potential 
via 
\begin{align}
	\theta_n = - a g A^{1}_{n}. 
\end{align}
Their conjugate variables $L_n$, with $\left[\theta_n,L_m\right] = i \delta_{nm}$, are related to the electric field by
\begin{align}
	g L_n =  E_n.
\end{align}

Since $\theta_n$ is an angular variable, $L_n$ will have integer charge eigenvalues $p_n \in \mathbb{Z}$. Therefore, the local
Hilbert space spanned by the corresponding eigenvectors $\left| p_n\right\rangle$ is infinite, and $e^{\pm i\theta_n}$ are the ladder operators
\begin{align}
	e^{\pm i\theta_n} \left| p_n \right\rangle = \left| p_n \pm 1 \right\rangle.
\end{align}
The lattice equivalent of Gauss' law then reads
\begin{align}
	L_n - L_{n-1} = \phi_{n}^{\dagger} \phi_n - \frac{1}{2} \left( 1 - \left(-1\right)^n\right),
	\end{align}
which means that excitations on odd and even sites create $\mp 1$ units of flux, corresponding to ``electron'' and ``positron'' excitations, respectively.
Using a Jordan-Wigner transformation, $\phi_n = \Pi_{k<n} \left(i \sigma_{k}^{z} \right) \sigma_{n}^{-} $, where $\sigma^{\pm}= \frac{1}{2} \left(\sigma^{x} \pm \sigma^{y}\right)$, the fermionic degrees of freedom can be mapped to spin-1/2 degrees of freedom while keeping the Hamiltonian local, i.e., 
\beqa
	\label{eq: Hamilton_without_gauge}
	H &=& \frac{g}{2\sqrt{x}} \Bigg( \sum_{n} L_{n}^2  + \frac{\mu}{2} \sum_{n } \left(-1\right)^n \left( \sigma_{n}^{z} + \left(-1\right)^n \right)  \nonumber \\   
       &+& x \sum_{n} \left(\sigma_{n}^{+} e^{i \theta_n}\sigma_{n+1}^{-} \right) + h.c. \Bigg).
       \label{eq:citeme}Ê
\eeqa
In the above equation we introduced the parameters $x \equiv 1/\left(g^2 a^2\right)$ and $\mu \equiv 2 \sqrt{x} m/g$.
The spins live on the sites of the lattice, with $\sigma_n^{z}\left|s_n\right\rangle = s_n\left|s_n\right\rangle$, and represent ``positrons'' on even sites and ``electrons'' on odd sites. 
An even site with $s_{2n} = -1$ corresponds to an empty positron state, while $s_{2n} = 1$ represents an occupied positron state, and vice versa for the odd electron sites.

 In $(1+1)d$, Gauss' law can therefore be rewritten as
 \beq
 L_n - L_{n-1} = 1/2 \left(\sigma_{n}^{z} +\left(- 1\right)^n\right),
 \eeq
 and can in fact be used to remove the gauge degrees of freedom \cite{SchwingerDMRG}. The resulting lattice Hamiltonian is then
 \beqa
 	H &=& x \sum_{n} \left(\sigma_{n}^{+} \sigma_{n+1}^{-} + \sigma_{n}^{-} \sigma_{n+1}^{+} \right) + \frac{\mu}{2} \sum_{n} \left(1+\left(-1\right)^n \sigma_{n}^{z} \right) \nonumber\\
 	&+& \sum_{n} \left(l + \frac{1}{2}\sum_{k=0}^{n} \left((-1)^k + \sigma_{k}^{z} \right)\right)^2,
 	\label{eq:gaugetermnonlocal}
\eeqa
where $l$ is a possible external background charge. In this new formulation there are no gauge variables but, however, we pay the price of having a non-local, long-range interaction term in the Hamiltonian. 

\subsection{Chiral condensate in the continuum} 

Let us now revise the so-called \emph{chiral condensate}. Without attempting to go into detail,  we discuss two continuous symmetries of the Schwinger model of which one is broken after quantization. In this context, the chiral condensate arises as an order parameter. 

The Lagrangian density of the Schwinger model is invariant under global phase transformations of the Dirac field, i.e.,
\begin{align}
	\psi'  = e^{i \alpha } \psi \rightarrow \mathcal{L'} = \mathcal{L},
\end{align}
where $\alpha$ is a real constant. According to Noether's theorem (see, e.g., Ref.\cite{QFT}), there is a conserved current $j^{\mu}$ associated with every continuous symmetry. In this case the \emph{vector current}  
\begin{align}
	j^{\mu} = \overline{\psi}\gamma^{\mu}\psi
	\label{eq:veccur}
\end{align}
is conserved, i.e.,
\begin{align}
	\partial_{\mu} j^{\mu} = 0. 
\end{align}
This global $U(1)$ symmetry is known to hold in fermionic field theory models although, in principle, the vacuum state could spontaneously break it \cite{Chiral1}.

Let us now consider the case of massless ($m = 0$) fermions. Then the Lagrangian of the Schwinger model has another continuous symmetry, namely the so-called \emph{chiral symmetry}. This symmetry implies that the Lagrangian density is invariant if one transforms $\psi$ into $\psi'$ as
\begin{align}
\psi' = e^{i \alpha \gamma_5} \psi.
\end{align} 
In the above equation $\gamma_5 \equiv \gamma_0 \gamma_1 $ anti-commutes with $\gamma_{\mu}$ for $\mu = 1,2$ and $\alpha$ is again a real constant.
For example, in the Dirac representation the gamma matrices are given by \footnote{See, e.g., Ref.\cite{Chiral1}.} 
\begin{align}
	\gamma_0 =  \begin{pmatrix}
1 & 0 \\ 
 0 & -1  
\end{pmatrix}, \quad
\gamma_1 = 
\begin{pmatrix}
0  & 1 \\ 
-1 & 0
\end{pmatrix}, \quad
\gamma_5 \equiv \gamma_0 \gamma_1 =
\begin{pmatrix}
 0 & 1 \\ 
 1 & 0 
\end{pmatrix}.
\end{align}
The associated Noether current for this symmetry is the so-called \emph{axial-vector current} $j_{5}^{\mu}$ which is given by
\begin{align}
	j_{5}^{\mu} = \overline{\psi} \gamma^{\mu} \gamma_5 \psi.
\end{align}
While the vector current in Eq.(\ref{eq:veccur}) is conserved in the quantized theory, the axial-vector current is not. 
This non-conservation of the axial-vector current is called \emph{chiral anomaly} or \emph{axial anomaly}. The divergence of the axial-vector current reads
\begin{align}
	\partial_{\mu} j_{5}^{\mu} = \frac{g}{2\pi} \epsilon_{\mu \nu} F^{\mu \nu},
\end{align}
where $\epsilon_{\mu \nu}$ is the Levi-Civita symbol in two dimensions, see Refs.\cite{Chiral1,Chiral2}.
As a consequence of this chiral symmetry breaking, the vacuum expectation value
\begin{align}
	\Sigma\equiv\left\langle \overline{\psi} \psi \right\rangle
\end{align}
becomes non-zero. The quantity $\Sigma$ is called \emph{chiral condensate} \cite{Chiral1}. In the case of the massless Schwinger model, the chiral condensate can be computed exactly (see, e.g., Ref.\cite{Wipf}),  and is found to be 
\begin{align}
	\Sigma_0 = \frac{e^{\gamma}}{2 \pi^{\frac{3}{2}}} \approx 0.159929,
\end{align}
where $\gamma$ is the Euler-Mascheroni constant. Therefore, the chiral condensate can be regarded as an order parameter signalling chiral symmetry breaking in the vacuum.

\subsection{Chiral condensate on the lattice}

In the lattice formulation of the massive Schwinger model, it is possible to write the chiral condensate in terms of Pauli spin operators. It is easy to see that this reads
\begin{align}
	\Sigma\left(x\right) = \frac{\sqrt{x}}{N} \sum_{n} \left(-1\right)^n \left\langle \frac{1+\sigma_{n}^{z}}{2} \right\rangle,
\end{align}
where the expectation value is computed in the ground state and $N$ is the number of lattice sites. The naively-computed chiral condensate is known to be UV-divergent. In particular, it diverges logarithmically in the continuum limit, $a\rightarrow 0$.  
It has been argued that this divergence comes solely from the free theory at $g=0$. 
In the free case the chiral condensate on the lattice $\Sigma_{free} \left(x\right)$ can be computed exactly as
\begin{align}
	\Sigma_{free} \left(x\right) = - \frac{m}{\pi g} \frac{1}{\sqrt{1 + \frac{m^2}{g^2 x}}} K\left(\frac{1}{1 + \frac{m^2}{g^2 x}}\right), 
\end{align}
where $K\left(z\right)$ is the complete elliptic integral of the first kind. 
This result can be used to subtract the divergence from the computed chiral condensate in the interacting theory, and therefore to
renormalize it. In other words, we can define a so-called \emph{subtracted chiral condensate} $\Sigma_{sub}$, which allows for a continuum extrapolation, by 
\begin{align}
	\Sigma_{sub} = \Sigma\left(x\right) - \Sigma_{free}\left(x\right),
\end{align}
where $ \Sigma\left(x\right)$ denotes the computed chiral condensate. Details on the extrapolation procedure for numerical data will be given in the next chapters.

\section{Infinite DMRG}
\label{sec3}

Here we review the basics of the DMRG algorithm for infinite systems. Several formulations of this algorithm have been proposed in the literature. The approach taken here is similar to that in the second paper of Ref.\cite{idmrg}, where we consider both the case of one-site and two-site updates. For the sake of completeness, we also briefly review some of the basics of variational optimization algorithms over tensor networks \cite{tn}.

\subsection{MPS variational optimization}Ê

In general terms, we want to approximate the ground state of a Hamilitonian expressed as a Matrix Product Operator (MPO) by minimizing 
\begin{align}
	 E\left(\left|\psi\right\rangle\right) = \frac{\left\langle \psi \right| H\left|\psi\right\rangle}{\left\langle \psi \right. \left|\psi\right\rangle}
\end{align}
over the family of Matrix Product States (MPS) with bond dimension $\chi$. This can be achieved by introducing a Lagrange multiplier $\lambda$ that enforces normalization, so that the minimization reads 
\begin{align}
	\min_{\left|\psi\right\rangle \in \text{MPS}} \left(\left\langle \psi \right| H\left|\psi\right\rangle -\lambda\left\langle \psi \right. \left|\psi\right\rangle\right).
	\label{eq:varprob_ten}
\end{align}

\begin{figure}
	\centering
	\includegraphics[width=1.03\linewidth]{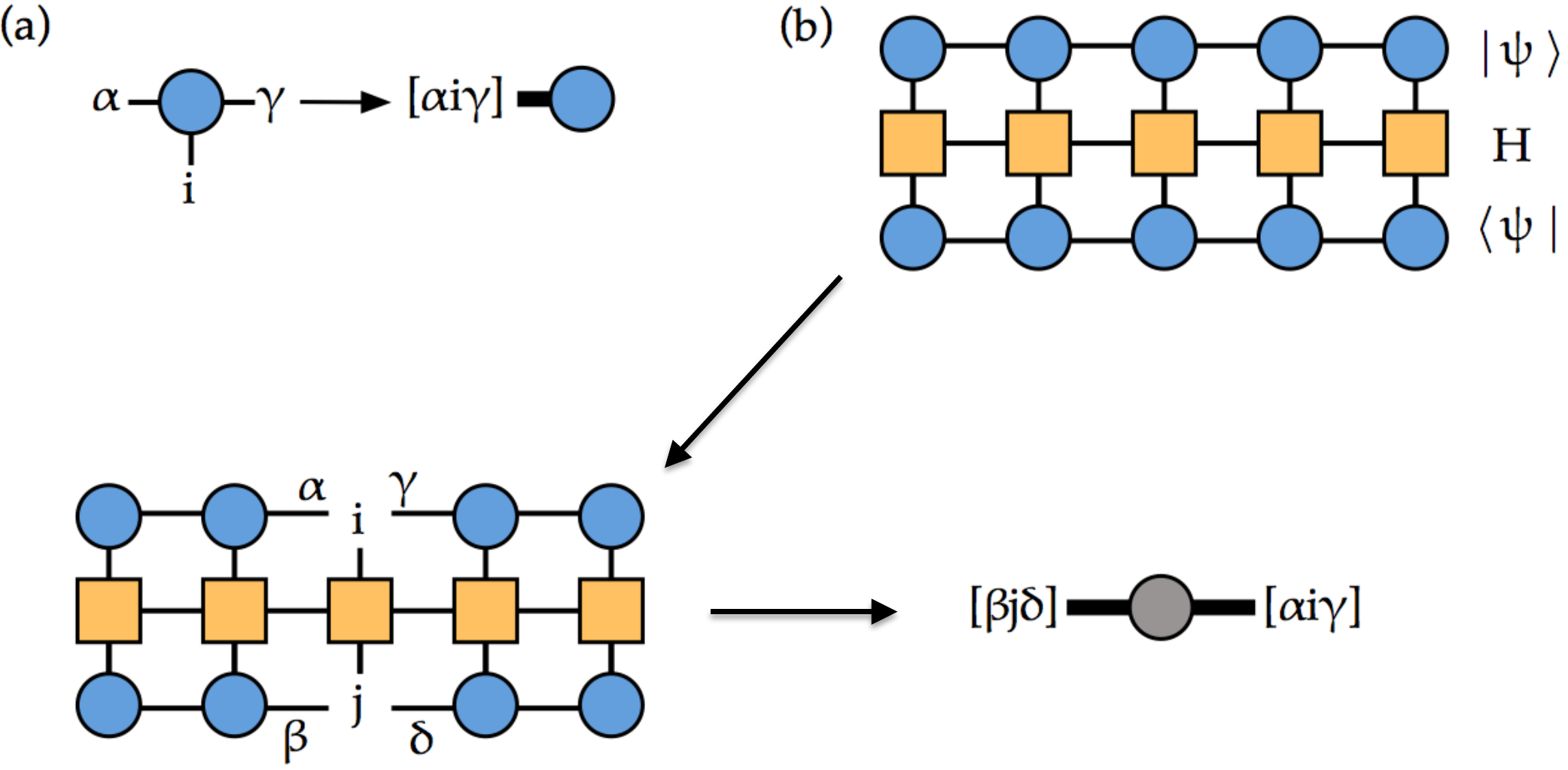}
	\caption{(Color online) (a) Transformation of a 3-index tensor into a vector by merging the indices;  (b)  Procedure to get the effective Hamiltonian for the third tensor in a 5-site MPS. }
	\label{fig:Heff}
\end{figure}

The above minimization is performed by adjusting all tensors in the MPS for all sites in order to make the expectation value of the energy the lowest possible. In DMRG one follows a sequential approach, optimizing tensor by tensor. In terms of the chosen tensor, which we call $A$, the mimization problem defined by Eq.(\ref{eq:varprob_ten}) can be written as
\begin{align}
\min_{A} \left(\left\langle \psi \right| H \left|\psi\right\rangle -\lambda\left\langle \psi \right. \left|\psi\right\rangle\right)
= \min_{A} \left( \vec{A}^{\dagger} \mathcal{H}_{eff} \vec{A}-\lambda \vec{A}^{\dagger} \mathcal{N} \vec{A}\right).
\end{align}
In the above equation, all coefficients of $A$ are arranged as a vector~$\vec{A}$ as shown in Fig.\ref{fig:Heff}(a), $\mathcal{H}_{eff}$ is an \emph{effective Hamiltonian}, and $\mathcal{N}$ is a \emph{normalization matrix}. The effective Hamiltonian and the normalization matrix can be considered as the $\emph{environment}$ of tensors $A$ and $A^{*}$ in the two TNs for $\left\langle \psi \right| H\left|\psi\right\rangle$ and $\left\langle \psi \right. \left|\psi\right\rangle$ respectively, but written in matrix form (see e.g. Fig.\ref{fig:Heff}(b)).

The minimization condition 
\begin{align}
	\frac{\partial}{\partial \vec{A}^{\dagger}}\left(\vec{A}^{\dagger} \mathcal{H}_{eff} \vec{A} - \lambda \vec{A}^{\dagger} \mathcal{N} \vec{A} \right) = 0
\end{align}
leads to the generalized eigenvalue problem
\begin{align}
 \mathcal{H}_{eff} \vec{A} = \lambda \mathcal{N}\vec{A}.
 \label{eq:GenEig}
\end{align}
Once this optimization with respect to $A$ is done, one proceeds by repeating the minimization for another tensor in the MPS. In this way, one continues sweeping through all tensors several times, until the desired convergence in expectation values is attained. Let us remark that if we start from an MPS with open boundary conditions, this algorithm is nothing else but the Density Matrix Renormalization Group (DMRG) algorithm in the language of TNs \cite{dmrg, tn}. In the case of open boundary conditions it is also always possible to choose an appropriate gauge for the tensors, e.g., a mixed canonical form with $A$ as the center site, such that $\mathcal{N} = \mathbb{I}$. Then Eq.(\ref{eq:GenEig}) reduces to an ordinary eigenvalue problem. This is very useful for practical implementations since it avoids stability problems due to $\mathcal{N}$ being ill-conditioned, see Ref.\cite{tn}. In what follows, we always consider MPS with open boundary conditions in mixed canonical form.
\begin{figure}
	\centering
	\includegraphics[width=1.03\linewidth]{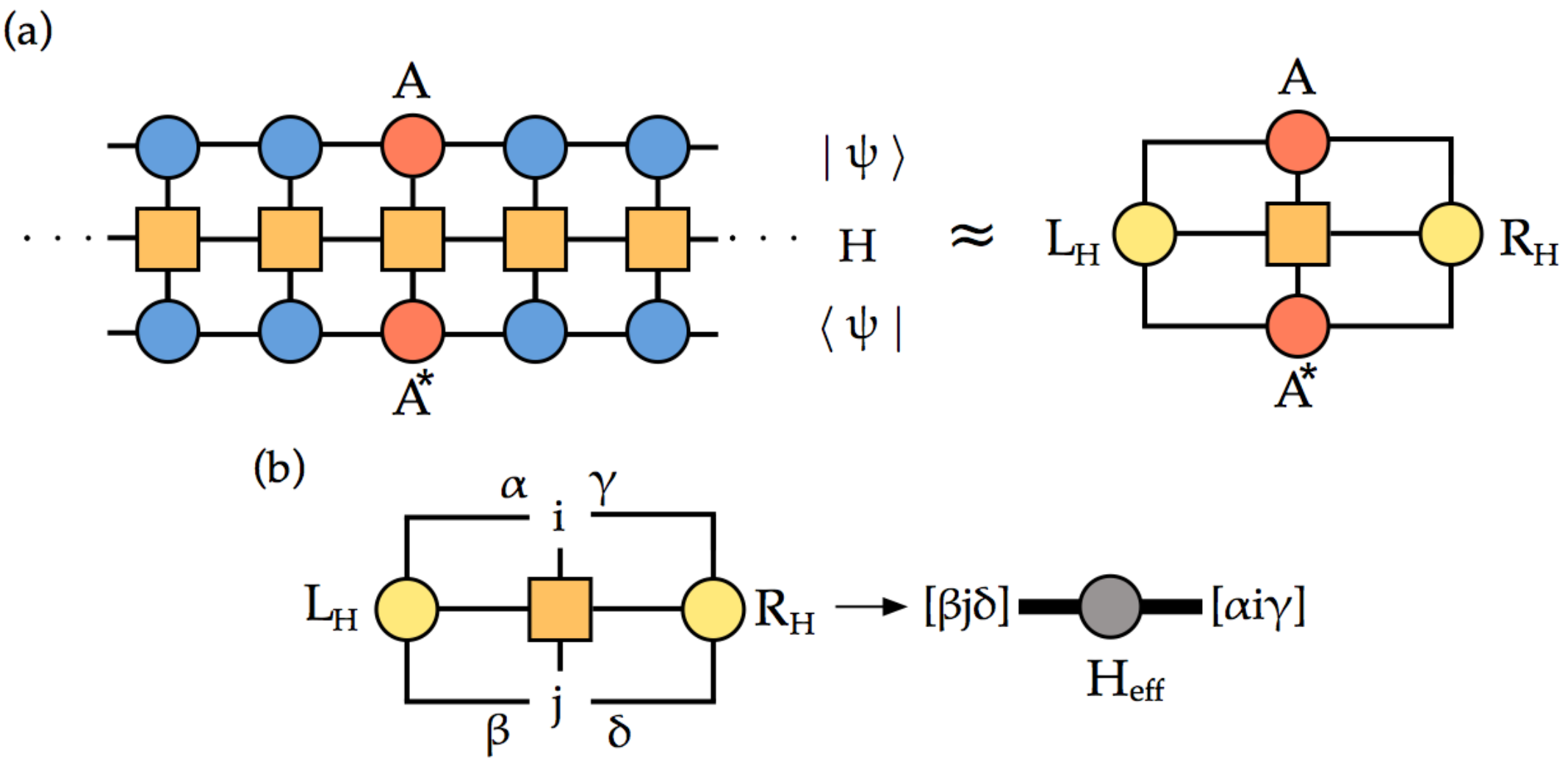}
	\caption{(Color online) (a) Definition of the approximate environments $L_H$ and $R_H$;  (b)  Definition of the effective Hamiltonian.}
	\label{fig:iDMRG1}
\end{figure}

\subsection{One-site infinite DMRG}
\label{s:iDMRG}
If we start from the very beginning with an infinite system to study systems in the thermodynamic limit, we need to modify the above procedure \cite{idmrg}. Let us assume that we were given an infinitely large and translationally invariant system in its ground state. Then, if we were to add an additional site to the system and allow it to relax, one would expect that the new site would change to match the rest, while the other sites in the system remain essentially unchanged. 
In MPS language, let us consider the case in which we already have an infinite MPS with bond dimension $\chi$ representing the ground state of our system. Then adding a site to our system would correspond to adding another tensor in the MPS. The relaxation process could be simulated by minimizing the energy with respect to the new tensor in the environment given by the original MPS. We would then obtain a tensor which looks, mostly, like all of the tensors in our infinite MPS. The idea of the algorithm is to start with a representation of the infinite system in terms of an \emph{approximative environment}. This environment is then progressively refined by embedding new sites, allowing the sites to relax, and then absorbing them into the environment. Eventually this procedure will converge, thus simulating the environment experienced by a single site in the infinite system in its ground state. 

\begin{figure}[btp]
	\centering
	\includegraphics[width=1.02\linewidth]{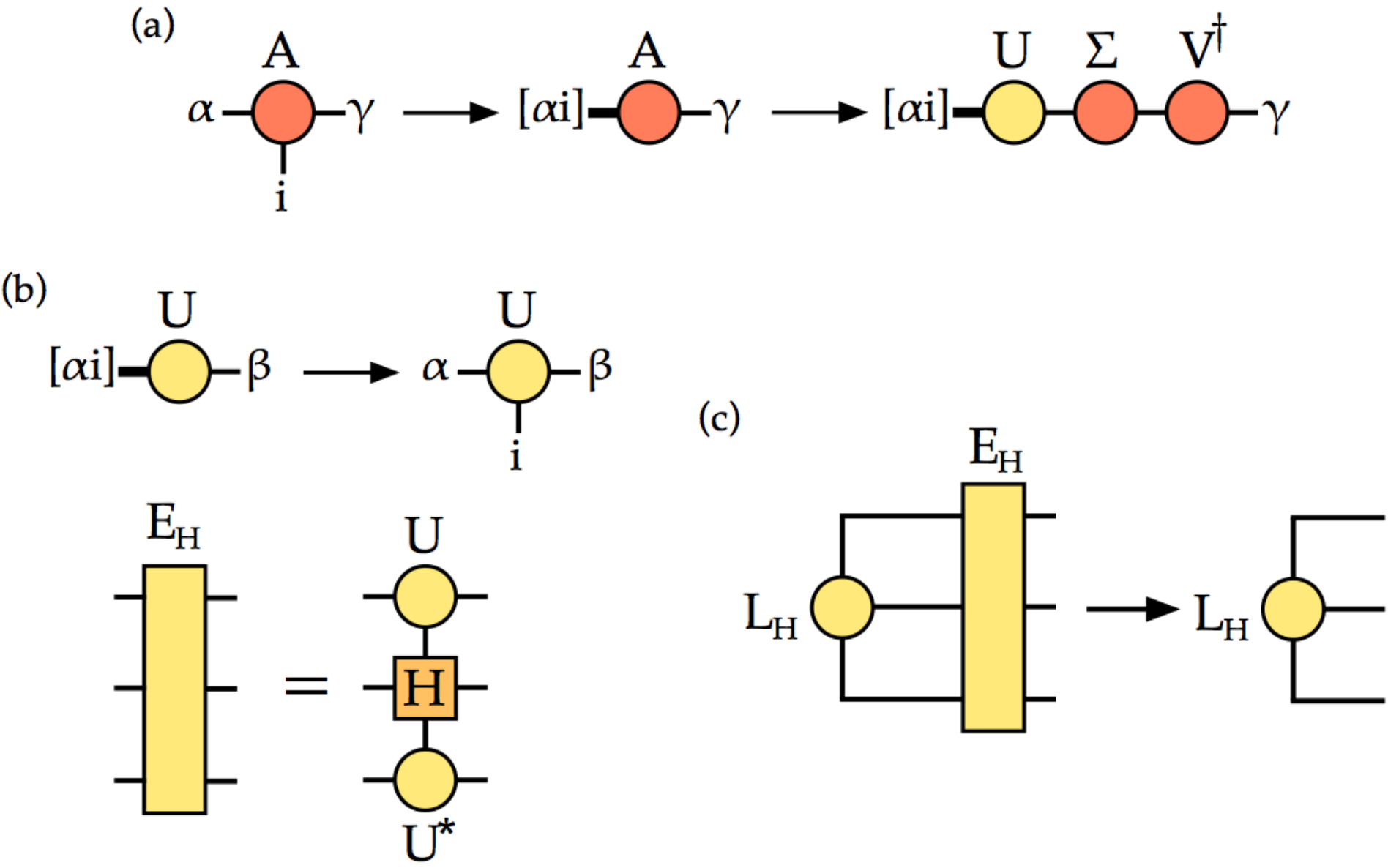}
	\caption{(Color online) Odd step: (a) SVD of the optimized tensor $A$;  (b) Definition of $E_H$; (c) Refinement of the left environment.}
	\label{fig:iDMRG2}
\end{figure}

\begin{figure}[btp]
	\centering
	\includegraphics[width=1.05\linewidth]{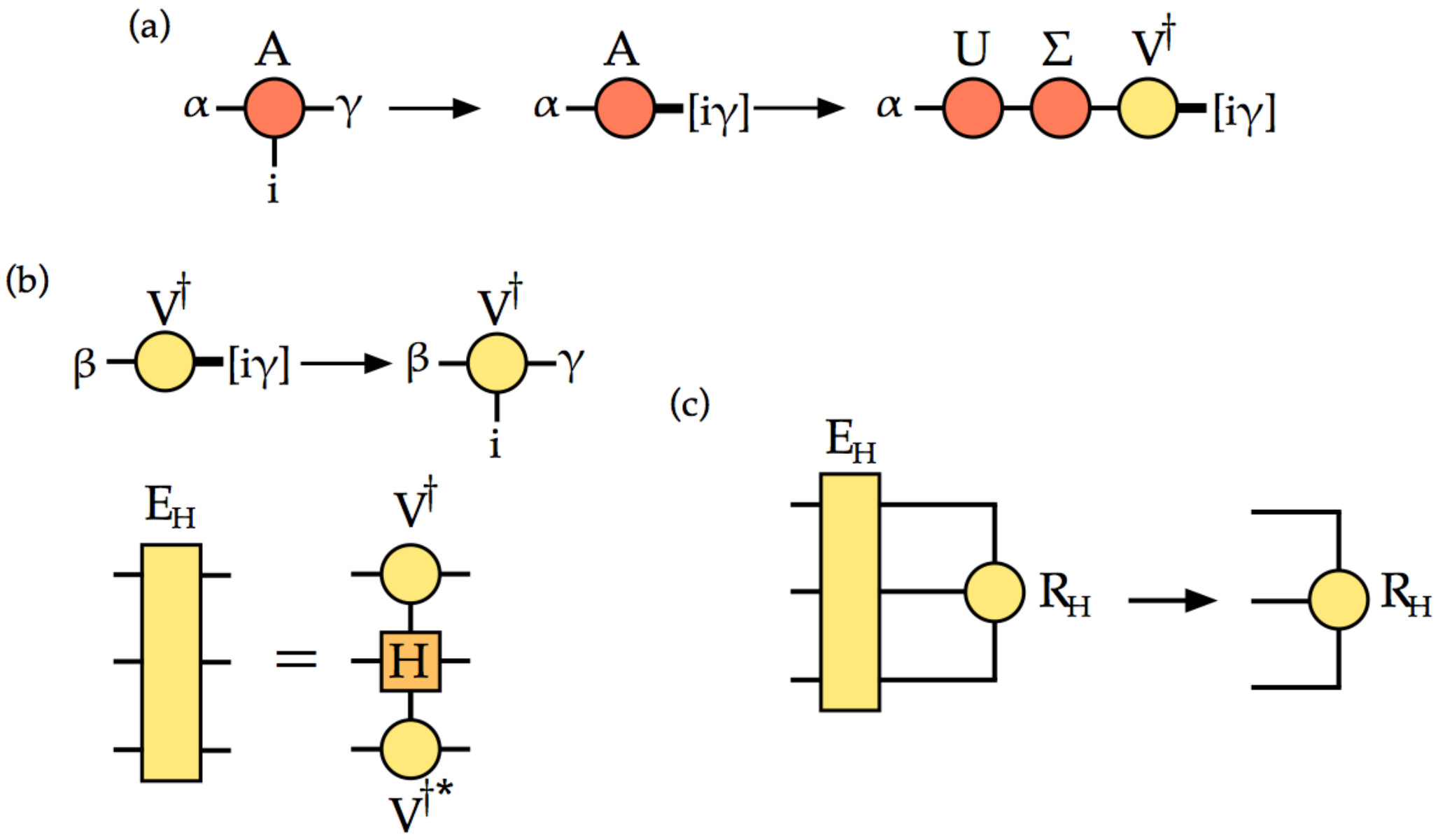}
	\caption{(Color online) Even step: (a) SVD of the optimized tensor $A$;  (b) Definition of $E_H$; (c) Refinement of the right environment.}
	\label{fig:iDMRG3}
\end{figure}

The infinite-system algorithm thus works as follows: one starts from initial environments $L_{H}$ and $R_{H}$ (e.g., random) representing the left and right halves of the (infinite) system with respect to the added tensor $A$ of the TN for $\left\langle \psi \right| H\left|\psi\right\rangle$ (see Fig.\ref{fig:iDMRG1}(a)). Then, one iterates the following procedure:  
\begin{enumerate}
	\item \emph{Relaxation:} compute the eigenvector $\vec{A}$ corresponding to the minimal eigenvalue of the problem \footnote{We choose $A$ as the center site for the mixed canonical form of the MPS.} $\mathcal{H}_{eff} \vec{A} = \lambda \vec{A}$, and reshape it back to a 3-index tensor. The effective Hamiltionian is shown in Fig.\ref{fig:iDMRG1}(b).
	\item \emph{Absorption (odd step)}: at an odd iteration step, the optimized tensor is absorved into the left environment $L_{H}$. In detail:
	\begin{enumerate}
	\item Merge the first bond index and the physical index of $A$ to form a matrix, and
	compute the singular value decompositon $A = U \Sigma V^{\dagger}$ (see Fig.\ref{fig:iDMRG2}(a)).  
	\item Undo the index fusion for the left index of $U$ to get back to a 3-index tensor (see Fig.\ref{fig:iDMRG2}(a))
	and compute $E_H$ as defined in Fig.\ref{fig:iDMRG2}(b).
	\item Refine the approximation for the left environment $L_H$ by contracting $E_H$ into it, i.e. $L_H \leftarrow L_H \cdot E_H$, as shown in Fig.\ref{fig:iDMRG2}(c). 
	\end{enumerate}	
	\item \emph{Absorption (even step)}: at an even iteration step, the optimized tensor is analogously contracted into the right environment $R_{H}$ (see Fig.\ref{fig:iDMRG3}). In detail:
	\begin{enumerate}
	\item Merge the second bond index and the physical index of $A$ to form a matrix, and
	compute the singular value decompositon $A = U \Sigma V^{\dagger}$.  
	\item Undo the index fusion for the right index of $V^{\dagger}$ to get back to a 3-index tensor  
	and compute the analogue of the tensor $E_H$. 
	\item Refine the approximation for the right environment $R_H$ by contracting $E_H$ into it, i.e. $R_H \leftarrow E_H \cdot R_H$. 
	\end{enumerate}	
\end{enumerate}
Since $U$ and $V$ are isometries, the mixed canonical form of the MPS is preserved at every simulation step. To check for convergence
it is useful to calculate the desired expectation value after, e.g., each first or second iteration step. For a single-site operator acting
on the added site this can be done easily, thanks specially to the mixed canonical form of the MPS. The main computational cost is given by the eigenvalue problem and scales therefore as $O\left(\chi^3\right)$.

\subsection{Two-site infinite DMRG}
\begin{figure}
	\centering
	\includegraphics[width=1.05\linewidth]{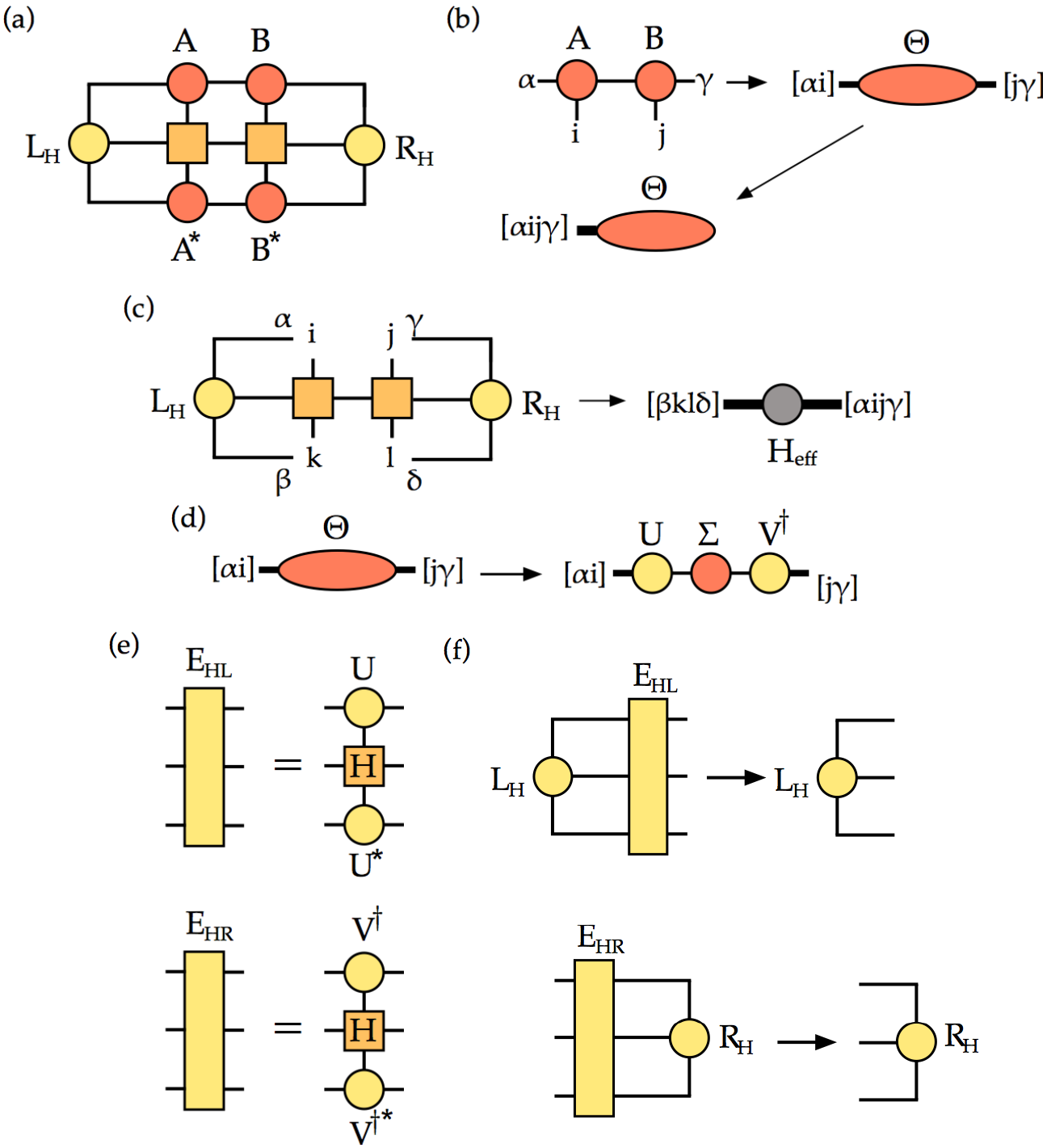}
	\caption{(Color online) Modifications for the two-site iDMRG algorithm: (a) 2-site environment; (b) Reshape of tensor $\Theta$ as a vector; (c) Reshape of the 2-site effective Hamiltonian; (d) SVD of tensor $\Theta$; (e) Tensors $E_{HL}$ and $E_{HR}$; (f) Left and right absorptions.}
	\label{fig:iDMRG4}
\end{figure}

If only a single site is added at every iteration step, then the MPS bond dimension $\chi$ is fixed right from the start in the algorithm. However, one may think of situations in which it would be advantageous to increase the bond dimension during the calculation. This can be done by a slight modification of the algorithm, namely, by adding \emph{two sites} at each iteration, see Fig.\ref{fig:iDMRG4}. The two-site infinite DMRG algorithm is then as follows:
\begin{enumerate}
	\item \emph{Relaxation:} compute the eigenvector $\vec{\Theta}$ corresponding to the minimal eigenvalue of the problem $\mathcal{H}_{eff} \vec{\Theta} = \lambda \vec{\Theta}$, where the effective Hamiltionian $\mathcal{H}_{eff}$
	and the vector $\vec{\Theta}$ are defined as shown in Fig.\ref{fig:iDMRG4}(c) and Fig.\ref{fig:iDMRG4}(b), respectively.
	\item \emph{Absorption}: the optimized tensor is simultaneously contracted into the left environment $L_H$ and into the right environment $R_H$. In detail:  
	\begin{enumerate}
		\item Compute the singular value decomposition $\Theta = U \Sigma V^{\dagger}$ (see Fig.\ref{fig:iDMRG4}(d))
		\item Undo the index fusion for the left index of $U$ and for the right index of $V^{\dagger}$.
		\item Compute the tensors $E_{HL}$ and $E_{HR}$ as defined in Fig.\ref{fig:iDMRG4}(e).
		\item Refine the approximations for the left environment $L_H$ and for right environment $R_H$ by the contractions $L_H \leftarrow L_H \cdot E_{HL} $ and $R_H \leftarrow E_{HR}\cdot R_{H}$ shown
		in Fig.\ref{fig:iDMRG4}(f).  
   \end{enumerate}
\end{enumerate}

The crucial point is that, if one adds two sites at a time, then the central matrix becomes a square matrix of increased dimension $d\chi \times d\chi$ as can be seen in Fig.\ref{fig:iDMRG4}(b). This allows, in principle, for a truncation of the SVD in the second iteration step, see also Fig.\ref{fig:iDMRG4}(d), which allows the bond dimension to grow as the algorithm proceeds. This is also particularly useful if one implements symmetries in the algorithm, since the truncation allows to change the symmetry sectors being kept at every step. In practice, this means that the algorithm can readapt itself to more relevant symmetry sectors, which have more weight in terms of the singular values of $\Theta$, leading to improved accuracy. 
      
\section{Gauge-invariant infinite DMRG}
\label{sec4}

Following the ansatz introduced in Ref.\cite{gaugeMPS}, we start one step before integrating out the gauge field degrees of freedom using the Gauss' law constraint. This is, we consider the Hamiltonian in Eq.(\ref{eq: Hamilton_without_gauge}), with spin variables on the sites for the staggered fermionic matter field, and angular variables on the links for the bosonic gauge (electric) field. An obvious advantage is that this Hamiltonian is \emph{local} with at most nearest-neighbour actions, and \emph{translationally invariant} under shifts by two sites. Furthermore, it is practical for possible generalizations to higher-dimensional systems, since the gauge degrees of freedom can only be integrated out in $(1+1)d$. 

 \subsection{MPO representation of $H$}
\begin{figure}
	\centering
	\includegraphics[width=1.1\linewidth]{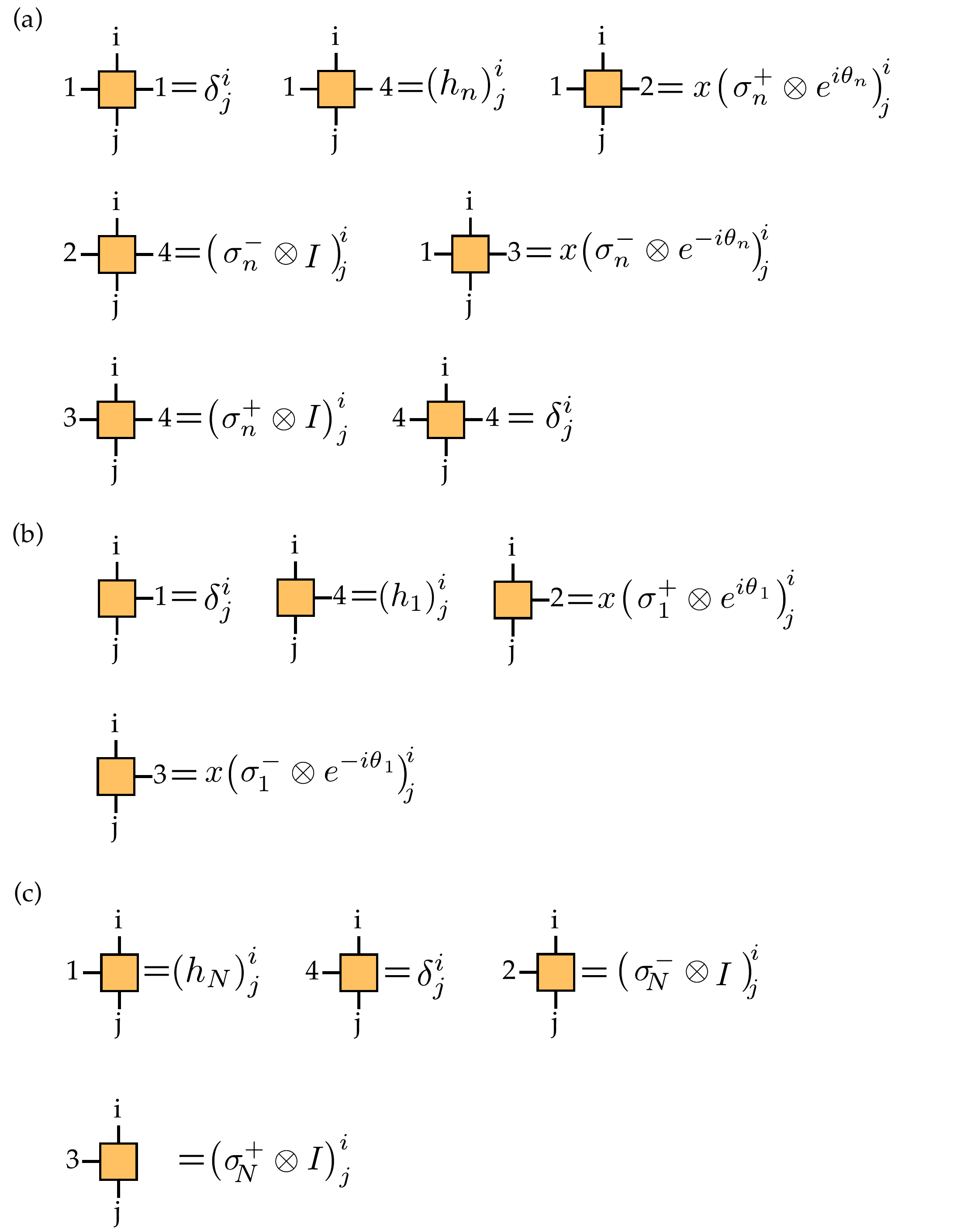}
	\caption{(Color online) MPO tensors for (a) the bulk, (b) the left boundary, and (c) the right boundary. Notice that we have different tensors for even and odd sites in the bulk due to the factor $\left(-1\right)^n$ in Eq.(\ref{eq:min}).}
	\label{fig:MPO_Hamilton}
\end{figure}

In the following, we provide an MPO representation of the Hamiltonian in Eq.(\ref{eq:citeme}) to be used in an iDMRG simulation. For convenience, we block site $n$ and link $n$ into a single MPS-site, such that at every MPS-site we have a fermionic and a gauge field degree of freedom. Then, the Hamiltonian can be regarded
as the sum of 1-site operator and 2-site operators, i.e, 
\begin{align}
	H = \sum_{n} h_n + h_{n,n+1}, 
\end{align}
where
\begin{align}
h_n = \mathbb{I} \otimes L_{n}^{2} + \frac{\mu}{2}\left(\left(\mathbb{I} + \left(-1\right)^n \sigma_{n}^{z}\right)\otimes \mathbb{I}\right),
\label{eq:min}
\end{align}
and
\beqa
h_{n,n+1} &=& x \left(\vphantom{\frac{1}{1}}\left(\sigma_{n}^{+} \otimes e^{i \theta_n}\right) \cdot \left(\sigma_{n+1}^{-}\otimes \mathbb{I}\right)\right. \nonumber \\Ê
 &+& \left.\left(\sigma_{n}^{-} \otimes e^{-i \theta_n}\right) \cdot \left(\sigma_{n+1}^{+} \otimes \mathbb{I}\right) \vphantom{\frac{1}{1}}\right). 
\eeqa
The first factor in the tensor product $\otimes$ refers to the fermion degree of freedom, and the second to the gauge field degree of freedom at the MPS site. With $\cdot$ we denote here the tensor 
 product between operators acting on different MPS-sites. The Hamiltonian can be written as an MPO with bond dimension $D = 4$ where non-zero coefficients of the tensors are given as in Fig.\ref{fig:MPO_Hamilton}. 

\subsection{Imposing gauge invariance}
We now impose gauge invariance to enforce that our algorithm  
works directly within the physical subspace of the full Hilbert space.   
In particular, we are only interested in states $\left| \psi \right\rangle$ that are gauge invariant, i.e., 
\begin{align}
	G_n \left| \psi \right\rangle = 0 \quad \forall n, 
	\label{eq:ConstraintGauss}
\end{align}
where 
\begin{align}
	G_n \equiv L_n - L_{n-1} - \frac{1}{2} \left(\sigma_{n}^{z}+\left(-1\right)^n\right).
	\label{eq:gl}
\end{align}
Eq.(\ref{eq:ConstraintGauss}) is nothing but the (discretized) lattice version of the Gauss' law constraint for the system. 

A possibility to impose gauge invariance would be to add a penalty term to the Hamiltonian, so that the gauge-invariant subspace is energetically preferred. For example, one could consider the modified Hamiltonian
\begin{align}
 H' = H + \lambda \sum_n G_{n}^{2},
\end{align}
instead of $H$, and then take the ground state sector in the limit $\lambda \rightarrow \infty $. However, by doing this gauge invariance would only be \emph{approximately} realized, and one would have to extrapolate additionally in parameter $\lambda$. 

\begin{figure}
	\centering
	\includegraphics[width=1\linewidth]{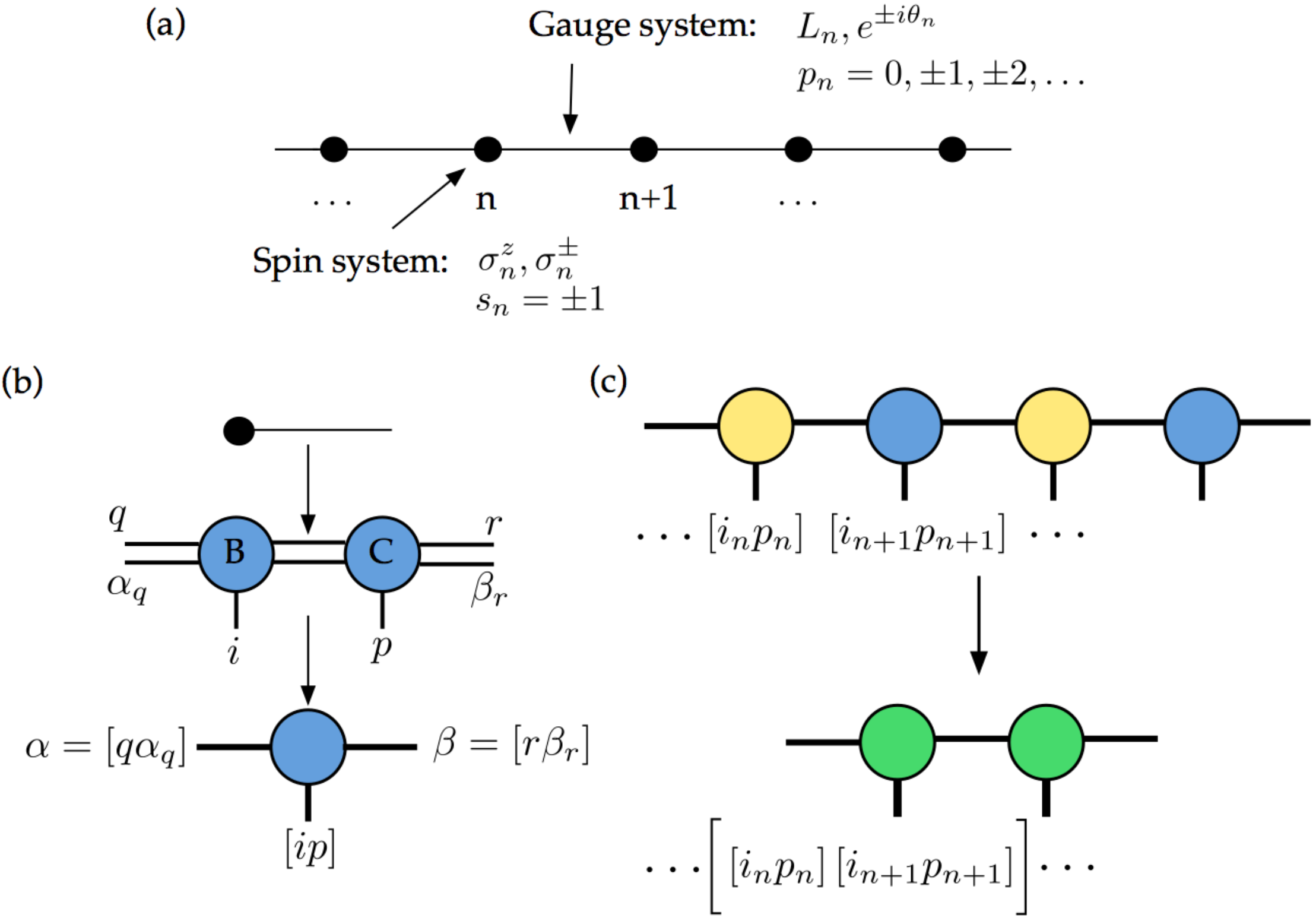}
	\caption{(Color online) (a) Infinite spatial lattice in $(1+1)d$: the spins (fermions) live on the sites and the gauge variables on the links; (b) A lattice site and the link to the right a represented by one MPS-site where gauge invariance is ensured by choosing the tensors as in Eq.(\ref{eq:MPSgauge}); (c) Neighbouring MPS-sites corresponding to positron and electron spin-gauge systems are blocked together in a ``supersite" to make the system fully translationally invariant.}
	\label{fig:Coarse}
\end{figure} 

A safer and more direct option is to implement gauge invariance directly at the level of the tensors in the TN, i.e., consider a TN made of $U(1)$ gauge-symmetric tensors \cite{gaugeMPS}. This implies that many tensor components in the MPS ansatz must vanish, i.e., only components compatible with gauge symmetry can be different from zero \cite{symTN}.  
  
For the sake of concreteness, let us assume that we have a finite lattice of $N\in 2\mathbb{N}$ sites. Then a general, i.e., not necessarily gauge invariant, MPS ansatz for the system has
 the form
 \beq
 \ket{\psi}Ê= \sum_{\{s_n, p_n \}} (B_{1})^{s_1}(C_{1})^{p_1}(B_{2})^{s_2}(C_{2})^{p_2}\cdots  \left| s_1 p_1 s_2 p_2 \ldots \right\rangle,
 \eeq
 where the matrices $(B_{n}^{s_n})_{\alpha \beta}$ correspond to fermionic degrees of freedom, and the matrices $(C_n^{p_n})_{\alpha \beta}$ to gauge degrees of freedom. 
We denote the bond dimension with $\chi$, i.e., the bond indices take the values $\alpha,\beta = 1,\ldots \chi$.

\begin{table}
	\centering
	\begin{tabular}{||c|c||} 
	\hline 
	~~~Parameter~~~ & ~~~~~~~Description~~~~~~ \\
         \hline 
         \hline Ê
	$\chi_{c}$ & Bond dimension of charge index \\
$\chi_{d}$ & ~~Bond dimension of degeneracy index ~\\
$p_{max}$ & ~~~Gauge boson truncation ~~\\
$N$ & Number of added sites \\
$x$ & Inverse coupling \\
$m/g$ & Dimensionless fermion mass \\ \hline
	\end{tabular}
	\caption{Simulation parameters in the one-site iDMRG algorithm.}
\label{tab:Simulation}
\end{table}

From Eq.(\ref{eq:ConstraintGauss}) and Eq.(\ref{eq:gl}) we can see that Gauss' law is basically a prescription to update the electric field $L_n$ at the right link of site $n$, namely,
\begin{align}
	L_n = L_{n-1} + \frac{1}{2} \left(\sigma_{n}^{z}+\left(-1\right)^n\right).
\end{align}
Therefore, if there is no charge at the site $n$, then $L_n$ stays with the value $L_{n-1}$ at the left. At the same time the electric field $L_n$ is increased/decreased by one unit if there is a positron/electron \footnote{Recall that an occupied positron or electron state corresponds to $s_n=1$ or $s_n= -1$, respectively, and that positrons/electrons live on even/odd sites $n$.} at site $n$. This ``update rule" can be implemented by giving the bond indices a multiple index structure, $\alpha\rightarrow \left(q,\alpha_q\right)$, and imposing the following form on the tensors in the bulk:
\begin{align}
	\label{eq:MPSgauge}
	&(B_n)^{s_n}_{\left(q,\alpha_q\right)\left(r,\beta_r\right)} = (b_{n,q})^{s_n}_{\alpha_q ,\beta_r} \delta_{q+\left(s_n+\left(-1\right)^{n}\right)/2,r}, \\ \nonumber
	&(C_n)^{p_n}_{\left(q,\alpha_q\right)\left(r,\beta_r\right)} = (c_{n})^{p_n}_{\alpha_q ,\beta_r} \delta_{q,p_n} \delta_{r,p_n}. 
\end{align} 
If one chooses a vanishing electric field to the left of the first lattice site, i.e. $L_0 = 0$, then the tensors representing the boundaries are gauge invariant if:
\begin{align}
	\label{eq:bound}
	&(B_1)^{s_1}_{\left(q,\alpha_q\right)\left(r,\beta_r\right)} = (b_{1,0})^{s_1}_{1,\beta_r} \delta_{\left(s-1\right))/2,r}, \\ \nonumber
	&(C_{2N})^{p_{2N}}_{\left(q,\alpha_q\right)\left(r,\beta_r\right)} = (c_{2N})^{p_{2N}}_{\alpha_q ,1} \delta_{q,p_{2N}}. 
\end{align}
In the above equations, the indices $q$ and $r$ label the \emph{electric charge sector}, and are sometimes referred to as \emph{structural} or \emph{charge} indices. They label the representation of the gauge symmetry group
for the index, and run from $1$ to a structural bond dimension $\chi_c$. The indices $\alpha_q$ and $\beta_r$ label the \emph{degeneracy subspace} within each charge (symmetry) sector, and run from $1$ to a degeneracy bond dimension $\chi_d$. Every bulk or boundary tensor which is chosen according to Eq.(\ref{eq:MPSgauge}) or Eq.(\ref{eq:bound}), respectively, preserves the gauge symmetry \emph{exactly}. The variational freedom lies now within the matrices $b_{n,q}^{s_n}$ and $c_n^{p_n}$, and the total MPS bond dimension is given by $\chi = \chi_c \cdot \chi_d$. The rather lengthy derivation of the result can be found in Ref.\cite{gaugeMPS}. We also refer the reader to Ref.\cite{symTN} for details on the implementation of symmetries in TNs and its consequences. 

\subsection{Further details}

The strategy presented above is very general. For an iDMRG simulation, the MPO is itself gauge-invariant by construction. If the MPS ansatz is also gauge-invariant, then the whole algorithm preserves gauge symmetry at every iteration step, of course provided that the initial conditions for the left and right environments are also gauge-invariant. This initial condition for the environment tensors is very easy to impose. 

\begin{figure}
	\centering
	\includegraphics[width=0.97\linewidth]{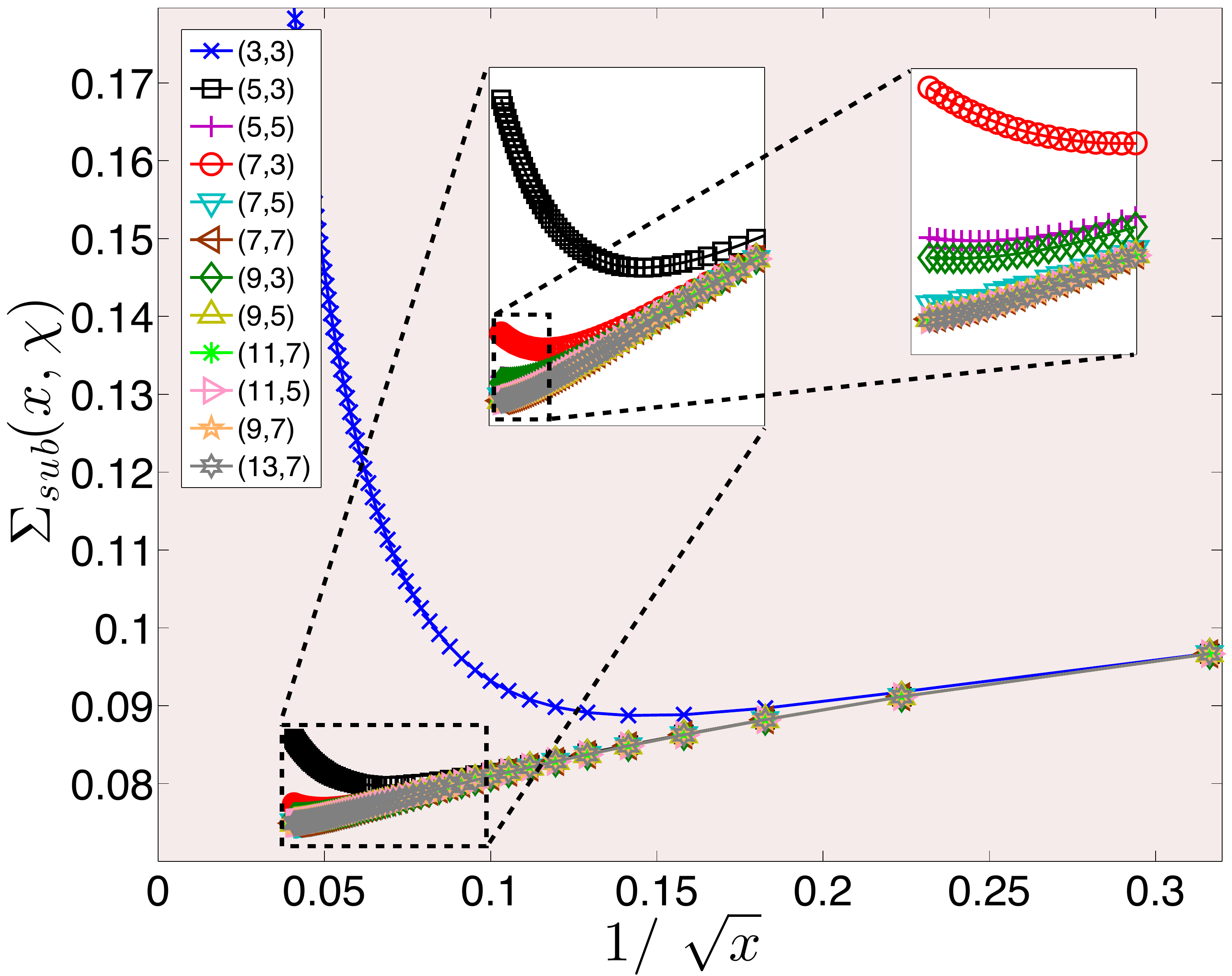}
	\caption{[Color online] Subtracted chiral condensate for $m/g = 0.25$ as a function of  $1/\sqrt{x}$ for different bond dimensions $(\chi_c, \chi_d)$, and physical gauge-boson dimension $5$. The insets show two consecutive zooms in the region with small lattice constant $a \sim 1/\sqrt{x}$.}
	\label{fig1}
\end{figure} 

Since our main goal is to learn from the simulations in $(1+1)d$, we use a coarse-grained version of the one-site iDMRG algorithm presented previously to find a ground state approximation in the thermodynamic limit (let us mention that we also tested a non-coarse-grained version of the two-site iDMRG algorithm, leading to essentially equivalent results). ÊAs in the construction of the Hamiltonian, we again block a lattice site and a link into one MPS-site. This leads to an MPS ansatz with a two-site unit cell due to alternating spin-gauge systems for positrons and electrons. The initial tensors
are defined according to Eq.(\ref{eq:MPSgauge}), but are otherwise chosen randomly (or according to some educated guess) within the variational gauge-invariant subspace. In order to obtain a system that is invariant under translations of one site, we also block neighbouring MPS-sites corresponding to a positron and electron spin-gauge systems together. This procedure is shown in Fig.\ref{fig:Coarse}.

 A list of all the relevant simulation parameters is shown in Table \ref{tab:Simulation}. 

\section{Results}Ê
\label{sec5}

\subsection{Numerical benchmarks}Ê

We computed the chiral condensate for four different values of the fermion mass, $m/g = 0, 0.125, 0.25, 0.5 $ where in each of the cases we took many points in the interval $x \in \left[10,600\right]$. Such a large interval allows us to extrapolate to the continuum limit, as well as to see the effect of the finite bond dimension as this limit is approached. The parameter $p_{max}\geq \left| p_n\right|$ truncates the \emph{infinite} local Hilbert space of the gauge bosons, and amounts to a maximum bosonic occupation number. Physically it can also be seen as truncation in the gauge $U(1)$ charge. In our calculations we choose $p_{max} = 2$, i.e., we truncate the infinite dimensional Hilbert space to five dimensions \footnote{$L_n \left| p_n \right\rangle = p_n \left| p_n \right\rangle$ with $p_n = 0, \pm 1, \pm 2$.}. In practice we have seen that this truncation is sufficient for our purposes \footnote{A more detailed analysis and justification of this truncation can be found in Ref.\cite{SchwingerTruncation}.}. Furthermore, we set $N = 500$ which corresponds to adding 1000 sites in the physical system due to the two-site coarse-graining. At every simulation step we check that the expectation value of the gauge operator $G_n$ defined in Eq.(\ref{eq:gl}) is zero, as required by gauge invariance.

\begin{figure}
	\centering
	\includegraphics[width=1\linewidth]{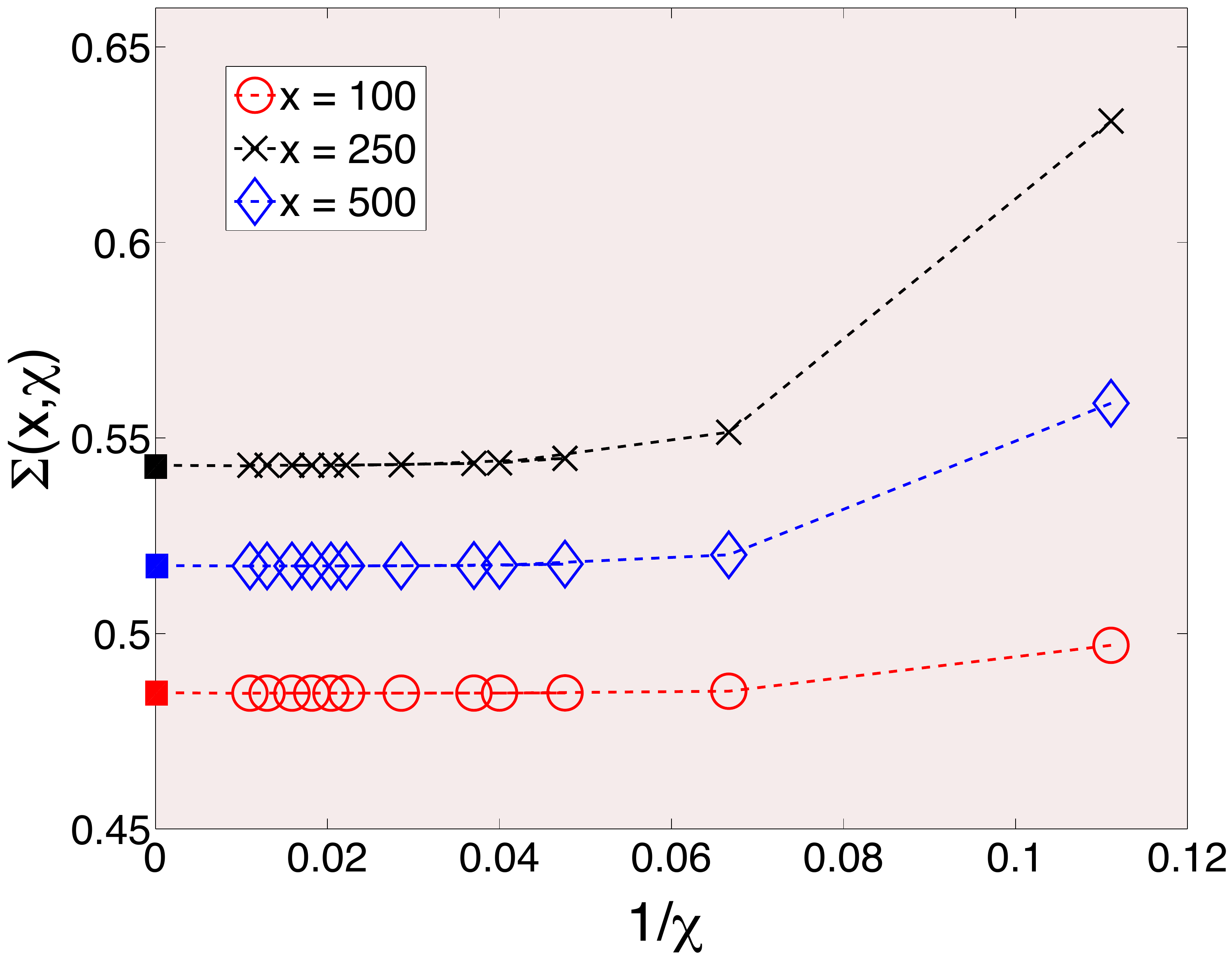}
	\caption{[Color online] Extrapolation of the computed chiral condensate in $1/\chi$, for $x=100, 250, 500$, at $m/g = 0.25$.}
	\label{fig2}
\end{figure} 

Importantly, in order to get an approximation of the (subtracted) chiral condensate in the continuum, we have to perform a sequence of  extrapolations. First, for every $x$ we extrapolate to infinite MPS bond dimensions $\chi_c$ and $\chi_d$. Second, for the extrapolated curve as a function of $x$, we extrapolate to the continuum limit so that $1/\sqrt{x} \rightarrow 0$. 

Let us show an example of the extrapolation in the MPS bond dimensions for $m/g = 0.25$. In Fig.\ref{fig1} we show the substracted  chiral condensate as a function of $1/\sqrt{x}$ for different bond dimensions $\chi_c$ and $\chi_d$. One can see that the effect of the truncation becomes stronger as the lattice parameter $a$ becomes smaller, i.e., in the region tending towards the continuum limit. This is an important observation: \emph{the closer we are to the continuum limit, the harder the simulation becomes}. It may be possible to simulate the lattice system always in an ``easy" regime far from the field theory limit, but it is important to remember that in such a case we would not be simulating a field theory, but rather some (interesting but discrete) lattice spin model. In our simulations, the results for the substracted chiral condensate seem to be well converged over the chosen spectrum of bond dimensions. 

In practice, for every $x$ we do an extrapolation in the total MPS bond dimension, i.e., $\chi \equiv \chi_c \cdot \chi_d \rightarrow \infty$. We find that the dependency of the chiral condensate is well described by the fitting function 
\beq
	 \Sigma\left(x, \chi\right) \approx a e^{-b\chi}+\frac{c}{\chi^d}+ \Sigma\left(x,\chi = \infty \right),
\eeq
where $\Sigma\left(x,\chi = \infty \right)$ is the value extrapolated to infinite bond dimension for inverse coupling $x$. in Fig.\ref{fig2}  we show some of these extrapolations for $x = 100, 250$ and $500$. 

\begin{figure}
	\centering
	\includegraphics[width=1\linewidth]{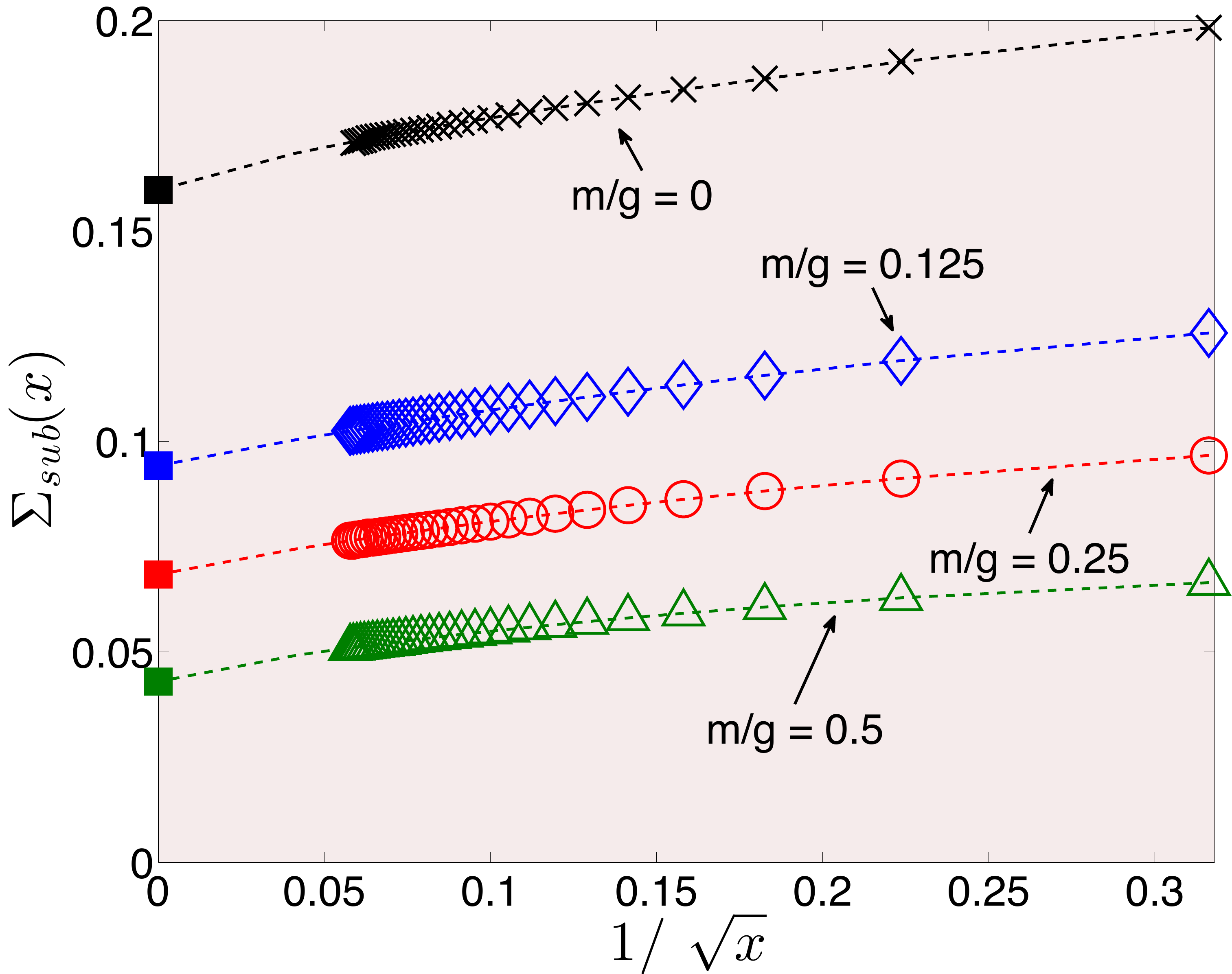}
	\caption{[Color online] Continuum extrapolation of the subtracted chiral condensate for $m/g = 0, 0.125, 0.25, 0.5$ attained from $x \in \left[10,300\right]$. Dashed lines correspond to the fit, and squares at $1/\sqrt{x} = 0$ to the extrapolated value in the continuum.}
	\label{fig3}
\end{figure} 

Finally, the extrapolation to the continuum limit $a\rightarrow 0$ is taken by considering the regime $x \rightarrow \infty$. Following the procedure in Ref.\cite{SchwingerDMRG}, we fit the subtracted chiral condensate using the following ansatz:
\begin{align}
	\Sigma_{sub}\left(x\right) = \Sigma_{sub} + F \frac{\log x}{\sqrt{x}} + B \frac{1}{\sqrt{x}} + C \frac{1}{x},
	\label{eq:fitansatz_extr}
\end{align}
where $\Sigma_{sub}$ is the extrapolated continuum value of the subtracted chiral condensate.
In Fig.\ref{fig3} we can see that this ansatz describes our data overall very well. This is especially true in the case of larger lattice constants, where the influence of the finite bond dimension is also smaller and the results are therefore easier to converge. Using this fit, we perform a continuum extrapolation for $x \in \left[10,300\right]$, where the convergence of our algorithm is particularly good. The obtained results for the four different fermion masses can be found in Table \ref{tab:results}. 

\begin{table}
	\centering
	\begin{tabular}{||c|c|c|c|c||} 
	\hline 
	$~m/g~$ & ~One-site iDMRG~ & ~Ref.\cite{SchwingerDMRG}~ & ~Ref.\cite{gaugeMPS} & exact \\
         \hline 
         \hline Ê
0     &  0.15900                                                                                   & 0.15993         & 0.15992                    & ~0.15992 \\ \hline
0.125 &  0.09425                                                                                   & 0.09202         & 0.09201                    & -        \\ \hline
0.25  &  0.06838                                                                                   & 0.06666        & 0.06664                    & -        \\ \hline
0.5   &  0.04293                                                                                   & 0.04238        & 0.04234                   & -        \\ \hline
0.75  &  -                                                                                         & -                   & 0.03062                     & -        \\ \hline
1     &  -                                                                                         & -                   & 0.02385                    & -        \\ \hline
2     &  -                                                                                         & -                   & 0.01246                   & -        \\ \hline
		\end{tabular}
\caption{Comparison: subtracted chiral condensate in the continuum. The extrapolation is in the regime $x \in [10, 300Ê]$.}
\label{tab:results}
\end{table}

\subsection{Discussion}Ê

As one can see in Table.\ref{tab:results} our results are in agreement with the results in Ref.\cite{SchwingerDMRG} and Ref.\cite{gaugeMPS} for a fitting region $x \in [0, 300]$, which is very well converged. Notice, though, that the approach in Ref.\cite{SchwingerDMRG} is conceptually very different, since it is based on \emph{finite-size} DMRG calculations using the \emph{non-local} Hamiltonian from Eq.(\ref{eq:gaugetermnonlocal}). In our gauge-invariant iDMRG approach, however, we start from the \emph{local} Hamiltonian in Eq.(\ref{eq: Hamilton_without_gauge}). We think that this approach is more convenient in order to generalize the calculations to higher-dimensional systems, since it preserves locality explicitly and is therefore more amenable to, e.g., approaches based on infinite Projected Entangled Pair States (iPEPS) \cite{iPEPS}. This is particularly true, also because in higher dimensions the Gauss' law cannot be integrated out, and therefore the most natural option is to work with a gauge-invariant 2d PEPS targeting a 2d local Hamiltonian on the lattice, as we shall discuss in Sec.\ref{sec4}.  

Let us also stress that in our $(1+1)d$ calculations with iDMRG, the bond dimensions did not need to be too large in order to get decent results. In particular, for $m/g = 0.25$ we used $\chi = 91$ as the highest total bond dimension, while in the other cases it was  $\chi = 63$. The extrapolations in Ref.\cite{SchwingerDMRG} were attained from calculations up to bond dimension $\chi = 140$, though via a different algorithm (as mentioned above). Remarkably, relatively small bond dimensions in iDMRG already allows us to provide results which are in quite good agreement with the ones for large bond dimension in Ref.\cite{SchwingerDMRG}, on top of not having to do any finite-size extrapolation since we work directly in the thermodynamic limit.  

For further comparison, in Ref.\cite{gaugeMPS}, besides working with gauge invariant MPS in the thermodynamic limit, the authors also exploited CT symmetry, i.e., invariance by a one-site translation and charge conjugation. The ground state calculations were done via the so-called \emph{time-dependent variational principle (TDVP)} \cite{tdvp}. In this work symmetries were treated in a more sophisticated way by distributing variational freedoms to different charge sectors. In contrast to that, our approach here is simpler, since we just choose gauge invariant initial tensors and then let the algorithm evolve, which naturally preserves gauge invariance. As such, it is remarkable that our simple approach produces results which are also in qualitative agreement with those produced by more sophisticated methods.  

Moreover, we remind that here we used the \emph{one-site} iDMRG algorithm. Despite being more efficient, we know that a two-site iDMRG calculation would bring some extra advantages, e.g., a dynamical increase of the bond dimension, and a dynamical truncation of the gauge-symmetry sectors. Still, we run some checks with a 2-site algorithm but did not obtain much greater accuracy in the regimes explored in this paper. However, it is good to keep in mind that the two-site approach may still be very useful in the more entangled regimes.

\section{Prospects for QED in $(2+1)d$}Ê
\label{sec6}

Taking into account what we have learned in the simulation of the $(1+1)d$ case, we would like now to consider the possibility of simulating the lattice version of QED in $(2+1)d$, directly in the thermodynamic limit. This gauge theory is interesting for a number of reasons: it is ``closer" to our $(3+1)d$ space-time and also has confinement \cite{confQED3} which, unlike in the case of the Schwinger model, appears through a mechanism much more similar to the one in $(3+1)d$ QCD \cite{QFT}. 

Simulating first the Schwinger model has allowed us to learn a number of useful things about how the simulation in $(2+1)d$ should proceed. In particular, for the $(2+1)d$ case one needs to face the following facts: 
\begin{enumerate}
\item{Gauss' law cannot be explicitly integrated out. Therefore, the safest choice is to work with a TN of gauge-invariant tensors.}
\item{The Jordan-Wigner transformation in $(2+1)d$ introduces non-local strings when mapping some fermionic terms into spins. Therefore, it is more convenient to work directly in fermionic Fock space.}
\item{Additionally to the electric field, in $(2+1)d$ there is a also magnetic field term which, in the lattice formulation, corresponds to a plaquette energy term in the Hamiltonian.}
\item{Moreover, and as in the Schwinger model, the gauge-boson Hilbert space should be truncated in a maximum occupation number in order to do the simulation (quantum link model) \cite{qlm}.}
\end{enumerate}Ê

Considering the above, and following the intuition built from the simulation of the $(1+1)d$ case, we would therefore need the following ingredients for the $(2+1)d$ simulation:  
\begin{enumerate}
\item{A TN in 2d as a variational ansatz in the thermodynamic limit. The so-called infinite-PEPS is the most natural option \cite{iPEPS}.}
\item{The ability to simulate fermions in 2d. This has already been achieved, with fermionic implementations of the iPEPS algorithm \cite{ftn}.}
\item{The ability to implement $U(1)$ gauge symmetry in the tensors. This has also been done already in 2d PEPS \cite{gtn, u1peps}.}
\item{The ability to deal with plaquette interactions. This has also been done in the past for iPEPS, e.g., when simulating the Toric Code model in a field and its generalizations \cite{plaqpeps}.}
\item{Efficient and accurate optimization strategies. Regarding this, important developments in 2d iPEPS methods have been put forward recently \cite{optpeps}.}
\end{enumerate}Ê

We conclude, therefore, that \emph{a priori} all the necessary ingredients for this simulation are already available. In the following section we would like to be a little bit more specific on how such a simulation could proceed. 

\subsection{Lattice formulation}
\begin{figure}
	\centering
	\includegraphics[width=0.35\linewidth]{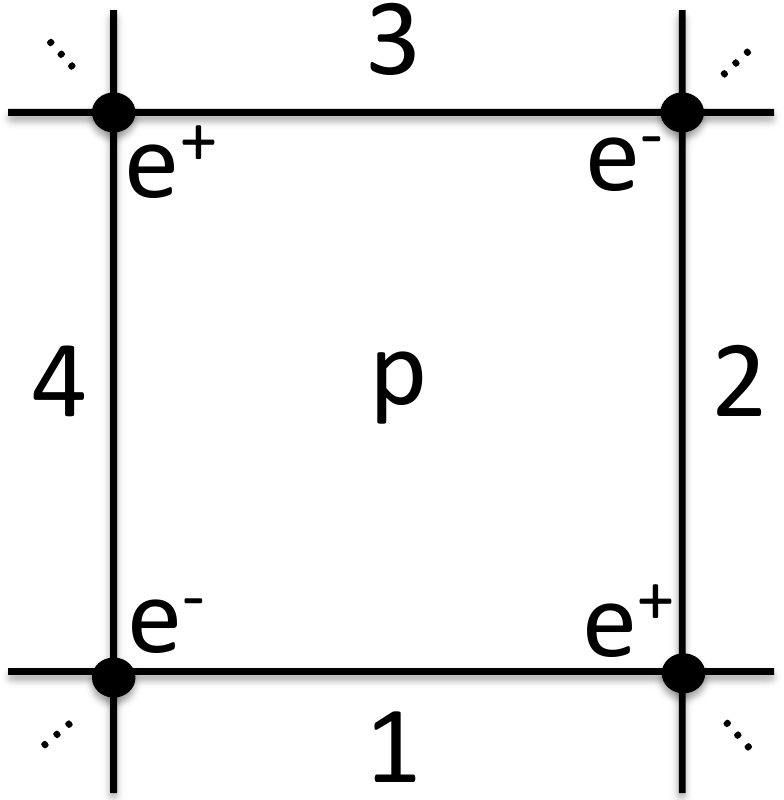}
	\caption{[Color online] Labelling of the links around a plaquette $p$, according to the magnetic term in the Hamiltonian of Eq.(\ref{2dqed}). The staggered structure of positrons $e^+$ and electrons $e^-$ is also shown.}
	\label{figqed3}
\end{figure}

Ê
The Hamiltonian of QED in $(2+1)d$ on a lattice can be derived in a similar way as the one in $(1+1)d$ in Sec.\ref{sec2}, but taking into account that this time one has two spatial dimensions instead of one. We give here a lattice Hamiltonian that has the correct continuum limit \footnote{See, e.g., Ref.\cite{dmrgOld} for more details.}. On a $2d$ spatial square lattice, the Hamiltonian is given by 
\beqa
H &=& - \frac{i}{2a}\sum_{\langle n, m \rangle} \left( \phi^\dagger_n e^{i \theta_{n,m}} \phi_m - h.c. \right) + m \sum_i (-1)^{s(n)} \phi^\dagger_n \phi_n \nonumber \\Ê
&+& \frac{ag^2}{2} \sum_{\langle n, m \rangle} L^2_{n,m} - \frac{1}{ag^2} \sum_p \cos{\left( \theta_1 + \theta_2 - \theta_3 - \theta_4 \right)}. 
\label{2dqed} 
\eeqa
In the equation above, $a$ is again the lattice spacing, $m$ the mass of the fermionic field, and $g$ the coupling between fermonic matter and the gauge boson. Fermionic fields are again staggered, but this time on a 2d square lattice, i.e., along both spatial directions, see Fig.\ref{figqed3}. The gauge boson variables $\theta_{n,m}$ live on the link between sites $n$ and $m$, and the sum $\langle n, m \rangle$ runs over nearest neighbours. The factor $s(n)$ decides the $+1$ or $-1$ prefactor for the mass term depending on the staggered pattern of the fermionic field: $+1$ for positrons, and $-1$ for electrons. Finally, the term with the cosinus is the curl of the gauge variable around a plaquette, see Fig.\ref{figqed3}, and corresponds therefore to the magnetic field energy. 

In this setting, the Gauss' law in $(2+1)d$ reads 
\beqa
L_{n,m} - L_{n,m+1} &=& \phi^\dagger_n\phi_n - \frac{1}{2} \left(1-(-1)^{s(n)} \right) \nonumber \\ 
L_{n,m} - L_{n+1,m} &=& \phi^\dagger_m\phi_m - \frac{1}{2} \left(1-(-1)^{s(m)} \right), 
\eeqa
where the first equation is for horizontal links, and the second for the vertical. Finally, in order to implement a simulation, it is advisable to truncate again the local dimension of the Hilbert space of the gauge boson, as we did in the $(1+1)d$ case. 

\subsection{Variational ansatz: a proposal}

\begin{figure}
	\centering
	\includegraphics[width=1.\linewidth]{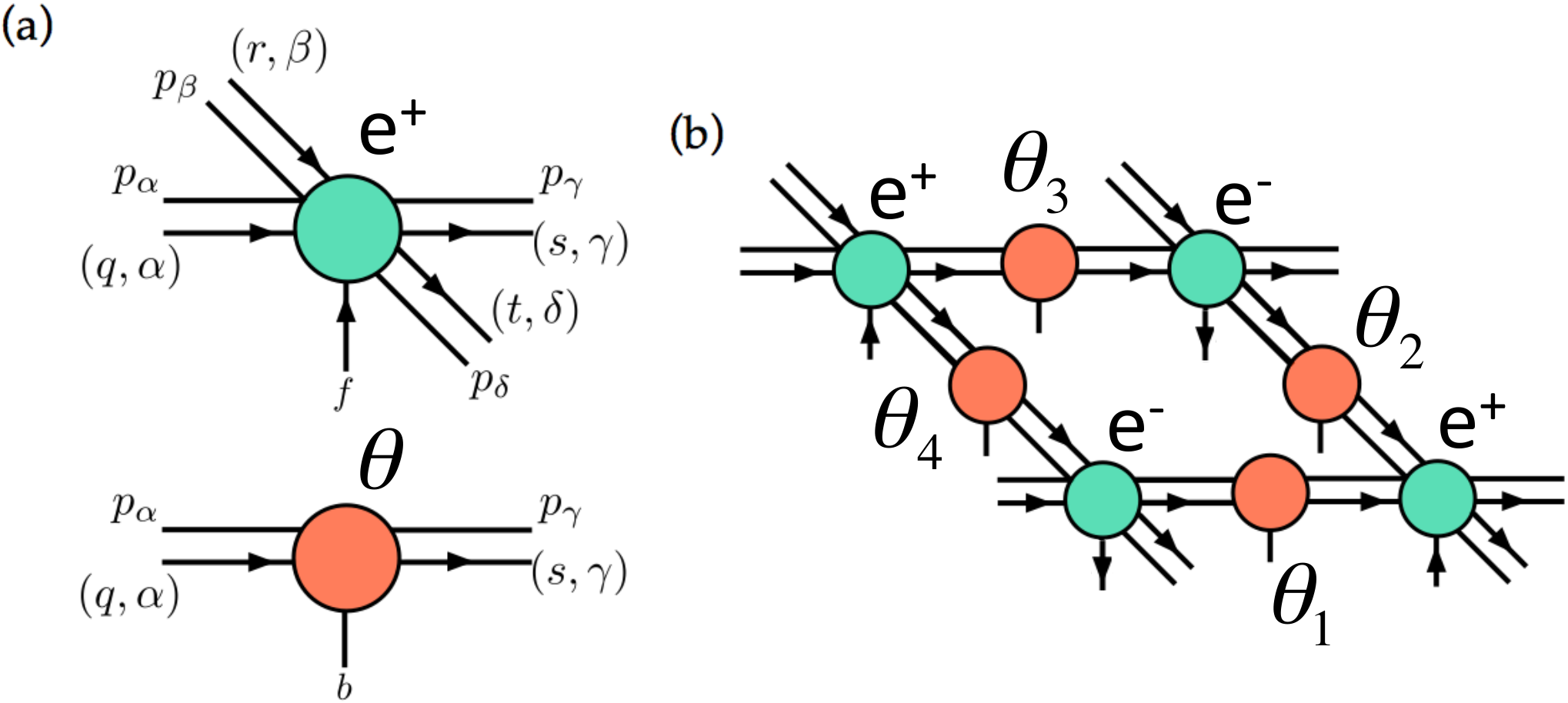}
	\caption{[Color online] PEPS variational ansatz for QED in $(2+1)d$. Tensors for fermionic variables are at the sites, and for bosonic gauge variables at the links; (a) The upper tensor is for a fermion, in fact a positron $e^+$. Its physical index is fermionic and oriented according to a $U(1)$-flux. Parity bond indices $p_\alpha, ..., p_\delta$ carry the fermionic parity, and are therefore fermionic and unoriented. Indices $(q,\alpha), ..., (t, \delta)$ carry the $U(1)$ charge and are bosonic and oriented. The lower tensor is for a gauge boson $\theta$. Its physical index is bosonic and unoriented. Its bond indices $p_\alpha, p_\gamma$ carry the fermionic parity, and are therefore fermionic and unoriented. Indices $(q,\alpha), (s, \gamma)$ are bosonic, carry the $U(1)$ charge, and are oriented; (b) Structure of a plaquette for the 2d infinite-PEPS. Notice the opposite orientation of the (fermionic) physical indices for positrons $e^+$ and electrons $e^-$, denoting their opposite $U(1)$ charges.}
	\label{qedpeps}
\end{figure}

As a variational TN ansatz to approximate the ground state of the above Hamiltonian we propose a 2d infinite PEPS with the structure from Fig.\ref{qedpeps}. There are two types of tensors: on the sites, for the staggered fermionic field (positrons and electrons), and on the links, for the bosonic gauge field. The physical indices at the sites are fermionic, as well as the unoriented bond indices. These indices satisfy the fermionic PEPS rules \cite{ftn}, namely, every time that two of such lines cross, one needs to include a fermionic swap gate in the TN diagram. Additionally, the physical indices at the links are purely bosonic and correspond to the truncated Hilbert space of the gauge variable for the corresponding link. Finally, bosonic bond indices are introduced with an orientation (arrow), which implement the $U(1)$ gauge symmetry in the tensor components.

In terms of equations, the non-zero components are the following for the tensors at the sites: 
\beqa
&&(B _\pm)^f_{(p_\alpha,(q,\alpha_q)),(p_\beta,(r,\beta_r)),(p_\gamma,(s,\gamma_s)),(p_\delta,(t,\delta_t))}  \\Ê
&=& b^f_{\alpha_q, \beta_r, \gamma_s, \delta_t} \delta_{mod(p_\alpha + p_\beta + p_\gamma + p_\delta + f,2),0} \delta_{(q + r \pm f), (s + t)}, \nonumber
\eeqa
where $\pm$ refers to a positron or an electron, tensor $b^f_{\alpha_q, \beta_r, \gamma_s, \delta_t}$ corresponds to the free parameters, the first delta implements fermionic $\mathbb{Z}_2$ parity symmetry, the second delta implements the gauge $U(1)$ symmetry, and $f = 0,1$ is the fermionic occupation number. Similarly, for the tensors at the link the non-zero components are given by
\beq
C^b_{(p_\alpha,(q,\alpha_q)),(p_\gamma,(s,\gamma_s))}  = c^b_{\alpha_q, \gamma_s} \delta_{mod(p_\alpha + p_\gamma, 2),0} \delta_{q,b} \delta_{s,b}, 
\eeq
where $c^b_{\alpha_q, \gamma_s}$ are the free variational parameters, $b$ is the bosonic physical index, the first delta implements the fermionic parity symmetry for the bond indices, and the last two deltas take into account $U(1)$ gauge symmetry. 

As mentioned above, this ansatz can be optimized in the thermodynamic limit to approximate the ground state of the Hamiltonian in Eq.(\ref{2dqed}). Such an optimization could be done variationally by using techniques recently introduced \cite{optpeps}, but it could also be optimized by imaginary time evolution with usual iPEPS algorithms \cite{iPEPS}. In any case, at every step in the algorithm one must carefully take into account (i) gauge invariance, as we did for the $(1+1)d$ case, but now also (ii) fermionic swaps, coming from the crossings of fermionic wires in the TN diagrams. The optimization of this ansatz by imaginary-time evolution is currently work in progress, and its results will be presented in a future publication. 

\subsection{Discussion} 

Let is now discuss briefly several aspects of QED in $(2+1)d$ that may be relevant for our simulation. First,  it is possible to consider compact and non-compact formulations of lattice QED, both with the correct continuum limit \cite{noncompact}. At the level of implementation, the main difference is the way we write the pure-gauge term, and both formulations on the lattice have slightly different behaviours for the scalings of the chiral condensate and the monopole density. In our case, non-compact QED in $(2+1)d$ could also be simulated with essentially the same scheme that we presented: in fact, one only would need to change the specific form of the plaquette gates. It would be interesting, thus, to benchmark both lattice formulations with our numerical approach. Second,  there is also the controversy about the dependence of  the chiral condensate with the number of flavours $N_f$ \cite{Nf}. Several works have argued in favour of a critical value $N_f^c$, so that there is no chiral symmetry  for $N_f > N_f^c$, though with no agreement on the actual value of $N_f^c$ and the type of phase transition. This is a problem that, in principle, could be explored also within our approach by including extra fermionic degrees of freedom for the flavours. However, this may involve larger bond dimensions in the ansatz, making the simulations more costly. Third, the interplay between fermions and monopoles is well known in compact QED in $(2+1)d$ \cite{mono}, where people have studied the possible survival of the  monopole plasma  in the presence of dynamical fermions, even in regimes where chiral symmetry is restored. This interesting question can also be addressed in principle by our method, studying the monopole density in terms of the number of flavours and the chiral condensate, up to the restrictions mentioned above. And fourth, there is also the issue of the finite-temperature dependence of the chiral condensate for QED in $(2+1)d$, with the presence of a confinement  - deconfinement transition conjectured to be of the BKT type \cite{fint}. Indeed, it should be possible to address this question with mixed-state versions of infinite-PEPS algorithms, which already exist in the literature for finite-temperature and even for dissipation \cite{finTTN}. In principle one could extend our variational ansatz to a PEPS-operator (PEPO) with the correct symmetries, to do a finite-temperature simulation.

\section{Conclusions} 
\label{sec7}

In this paper we have simulated the Schwinger model in the thermodynamic limit on a lattice, by using a gauge-invariant version of the iDMRG algorithm. After discussing the details of the theory and the particulars of one-site and two-site iDMRG, we have approximated the ground state and computed the extrapolation to the continuum of the substracted chiral condensate for several values of the coupling, in good agreement with alternative calculations. These results allowed us to build intuition on how a TN simulation of QED in higher dimensional systems should proceed. In particular, we proposed a gauge-invariant variational ansatz for the ground state of QED in $(2+1)d$ in terms of an infinite-PEPS with bosonic and fermionic degrees of freedom, as well as $U(1)$ gauge-invariant tensors. We discussed also that all the ingredients for such a simulation are in principle available in TN methods: $2d$ fermions, $U(1)$ gauge symmetry, plaquette interactions, and accurate optimization schemes. This simulation in $(2+1)d$ is currently work in progress. We hope that this paper will help to clarify, at least qualitatively, the ``big picture" towards TN simulations of lattice gauge theories in higher dimensions, with the target of lattice QCD in $(3+1)d$ on the horizon. We also hope that this paper helps to clarify, specially to the lattice gauge theory community, how one can handle the different ingredients of these field theories in the TN language directly in the thermodynamic limit, in order to simulate elusive regimes in quantum Monte Carlo.

\acknowledgements

We acknowledge discussions with M.-C. Ba$\tilde{{\rm n}}$uls, K. Cichy, I. Cirac, K. Jansen, E. Rico,  M. Rizzi and H. Saito, as well as the Donostia International Physics Center (DIPC), where part of this work was written.

 \end{document}